\documentclass[prd,aps,superscriptaddress,preprintnumbers,tightenlines,nofootinbib,eqsecnum,10pt,showpacs]{revtex4-2}
\usepackage{amsmath,amssymb,amsfonts,mathrsfs,bm,graphicx}
\usepackage{color}
\usepackage{orcidlink}
\usepackage[T1]{fontenc}
\usepackage[justification=raggedright]{caption}
\usepackage[caption=true]{subfig}
\usepackage{physics}
\usepackage{upgreek}
\usepackage{enumerate}
\usepackage{parskip}
\usepackage{tikz}
\definecolor{darkgreen}{rgb}{0.01, 0.75, 0.24}
\definecolor{darkblue}{rgb}{0.0, 0.2, 0.6}
\hypersetup{
	unicode=false,          
	pdftoolbar=true,        
	pdfmenubar=true,        
	pdffitwindow=false,     
	pdfstartview={FitH},    
	pdftitle={My title},    
	pdfauthor={Author},     
	pdfsubject={Subject},   
	pdfcreator={Creator},   
	pdfproducer={Producer}, 
	pdfkeywords={keyword1} {key2} {key3}, 
	pdfnewwindow=true,      
	colorlinks=true,       
	linkcolor=darkgreen,          
	citecolor=cyan,        
	filecolor=magenta,      
	urlcolor=darkgreen,           
	linktocpage=true
}

\allowdisplaybreaks
\renewcommand{\bfseries}{\fontfamily{ppl}\fontseries{bx}\selectfont}

\labelformat{section}{Section #1} 
\labelformat{subsection}{Section #1} 
\labelformat{subsubsection}{Section #1}
\labelformat{subsubsubsection}{Section #1}
\labelformat{equation}{Eq.~(#1)} 
\labelformat{figure}{Fig.~[#1]} 
\labelformat{subfigure}{Fig.~[\thefigure#1]} 
\labelformat{table}{Tab.~#1} 
\labelformat{appendix}{Appendix #1}
\usepackage{amsfonts}
\usepackage{amssymb}
\usepackage{physics}
\usepackage{bm}
\usepackage{mathrsfs}
\usepackage{graphicx}
\usepackage{enumitem}

\usepackage{upgreek}
\usepackage{ccfonts,palatino}
\hypersetup{
    colorlinks=true,
    linkcolor=darkblue,
    citecolor=cyan,
    urlcolor=darkgreen,
    pdfstartview={FitH},
    linktocpage=true
}

\allowdisplaybreaks

\DeclareSymbolFontAlphabet{\mathrsfs}{rsfs}
\DeclareMathAlphabet{\mathcal}{OMS}{cmsy}{m}{n}

\newcommand{\be}{\begin{equation}}
	\newcommand{\ee}{\end{equation}}
\newcommand{\bse}{\begin{subequations}}
	\newcommand{\ese}{\end{subequations}}
\newcommand{\ba}{\begin{align}}
	\newcommand{\ea}{\end{align}}

\makeatletter
\g@addto@macro\bfseries{\boldmath}
\makeatother
\usepackage{hyperref}
\usepackage{nameref}
\begin{document}
	\title{Photon Ring Dimming as a Signature of Photon-Axion Conversion in Janis-Newman-Winicour Naked Singularity}

	\author{Ayush Hazarika\orcidlink{0009-0004-5255-0730}}\email{ayush.hazarika4work@gmail.com}
	\affiliation{Department of Physics, Tezpur University, Napaam, Tezpur, 784028, Assam, India}
	\author{Premachand Mahapatra\orcidlink{0000-0002-3762-8147}}\email{p20210039@goa.bits-pilani.ac.in}
	\affiliation{Department of Physics, Birla Institute of Technology and Science-Pilani, K. K. Birla Goa Campus,\\NH-17B, Zuarinagar, Sancoale, Goa-403726, India}
	\author{Subhadip Sau\orcidlink{0000-0002-6271-1321}}\email{subhadipsau2@gmail.com (correspomding author)}
	\affiliation{Department of Physics, Jhargram Raj College, Jhargram, West Bengal-721507, India}
	\affiliation{Institute of Astronomy, Space and Earth Science (IASES), Bidhan Sishu Sarani, Kolkata-700054, India}
	\date{\today}
	
	\begin{abstract}
		
		The possible existence of axions in the universe introduces the intriguing possibility of photon-axion conversion in strong magnetic fields, particularly near compact objects like supermassive black holes or even naked singularity. In this study, we investigate the conversion of photons into axions in the vicinity of a Janis-Newman-Winicour (JNW) spacetime, a well-known naked singularity solution. We calculate the conversion probability and find that it is significantly influenced by the characteristic parameter of the JNW spacetime. The potential observational signatures of this conversion would be the dimming of photon ring in the X-ray and gamma-ray spectrum. Our findings suggest that compact objects like M87* could be prime candidates for detecting photon-axion conversion effects, provided future advances in high-resolution observations. Our analysis also suggests that the scattering of photons during the propagation through plasma is insignicant for the estimation of the conversion probability.
		
	\end{abstract}
	

	\maketitle

\newpage
\tableofcontents

\newpage
\section{Introduction}
Axions are hypothetical pseudoscalar elementary particle proposed as a solution to the strong CP (Charge Parity) problem in QCD (Quantum chromodynamics) \cite{peccei1977cp, weinberg1978new, wilczek1978problem, kim1979weak, shifman1980can, dine1981simple, zhitnitskii1980possible,Conlon:2006tq}.  { Photon-axion conversion has been thoroughly investigated for solar axions\cite{armengaud2014conceptual, cast2017new}, axion dark matter\cite{asztalos2010squid}, and various high-energy astrophysical phenomena\cite{csaki2002dimming, csaki2002effects, deffayet2002dimming, grossman2002effects}. The conversion of photons from axions in extragalactic space may influence gamma-ray\cite{galanti2023observability, Conlon:2014xsa,troitsky2022parameters, baktash2022interpretation, lin2023electroweak, gonzalez2023grb, nakagawa2023axion, carenza2022alp,wang2023axion} and X-ray emissions\cite{Conlon:2015uwa,Berg:2016ese,Conlon:2017qcw,hooper2007detecting, hochmuth2007effects, de2008axion, abramowski2013constraints, ajello2016search, marsh2017new, zhang2018new, reynolds2020astrophysical}, rendering these signals a potential investigation into axion physics. In cosmology, axions can have significant implications. For instance, heavy axions can achieve slow-roll inflation due to their shift symmetry \cite{freese1990natural,kim2005completing, dimopoulos2008n},  light axions are a potential dark matter candidate \cite{preskill1983cosmology, abbott1983cosmological,dine1983not,hui2017ultralight, chadha2022axion}. This has motivated extensive theoretical and experimental efforts aimed at detecting axions or axion-like particles (ALPs) through their coupling to photons, particularly in astrophysical settings where strong magnetic fields can induce photon-axion conversion.} Nevertheless, despite exploring various scenarios, the absence of observational evidence places limits on the axion coupling constant \cite{dolan2022advancing, dessert2022upper}.

{ 
In high-energy astrophysical settings, such as those adjacent to compact objects like black holes or neutron stars, intense magnetic fields facilitate the occurrence of photon-axion conversion. This coupling enables photons to oscillate into axions in a magnetic field, presenting a potential method for detecting axions through their indirect effects on electromagnetic signals. The effect is particularly significant in the X-ray and gamma-ray domains, where high-energy photons are more prone to conversion. Thus, examining the alteration of the electromagnetic spectrum in these settings has emerged as a valuable method for investigating the existence of axions and imposing limitations on their characteristics\cite{Nomura:2022zyy}.
}

{ Recent advancements in astrophysical observations, notably the detection of gravitational waves from binary mergers\cite{LIGOScientific:2016aoc,LIGOScientific:2017vwq} and the first image of the shadow of the supermassive black hole in M87* by the Event Horizon Telescope (EHT)\cite{Fish:2016jil,EventHorizonTelescope:2019dse,EventHorizonTelescope:2019uob,EventHorizonTelescope:2019jan,EventHorizonTelescope:2019ths,EventHorizonTelescope:2019pgp,EventHorizonTelescope:2019ggy}, have opened new opportunities for examining fundamental physics in strong gravity regime. The  successful imaging of the supermassive black hole M87* by EHT has yielded unparalleled insights into photon behavior near a  event horizon of a compact object, especially around the photon ring, a region where photons are transiently ensnared in near-orbit. This region, marked by strong gravitational lensing, provides a distinctive observational opportunity for exploring novel physics, including the possibility of axion-photon interactions. Such events may produce discernible marks on the photon ring, including frequency-dependent attenuation resulting from photon-axion conversion.
However, photon rings are not exclusive to black holes. Alternative compact objects, such as naked singularities, can also produce distinct photon rings due to the absence of an event horizon. In recent years, other horizonless compact objects, such as fuzzballs, have also been shown to produce radically different photon ring structures compared to black holes\cite{Mayerson:2023wck}. These investigations underscore the significance of photon rings as a probe for differentiating between various categories of compact objects, whether they are horizonless or have event horizons. {In recent study, the distinguishability of a naked singularity from a black hole has also been explored\cite{Dihingia:2024cch}}.
}

{
Although black holes have predominantly been the subject of research about photon rings and strong-field gravitational physics, alternative compact objects like naked singularities offer fascinating theoretical prospects. Naked singularities, devoid of an event horizon, are solutions to Einstein's field equations that challenge the cosmic censorship conjecture\cite{Penrose:1969pc}. The Janis-Newman-Winicour (JNW) naked singularity offers a clear spacetime framework to investigate the differences in photon behavior when an event horizon is absent compared to its presence near a black hole. The absence of a horizon modifies photon paths i.e geodesics, resulting in a photon ring around a naked singularity that may  exhibit unique characteristics,  that could be used to distinguish these objects from black holes.  In this context, we also wish to mention that  the possibility of the compact object at the core of our own Milkyway galaxy and also the at the core of Messier 87 be a naked singularity can not be ignored\cite{Lora-Clavijo:2023ukh,Pal:2023wqg,Mishra:2023uxl,Deliyski:2023gik,Nguyen:2023clb}.}

{

In this paper, we explore photon-axion conversion within the framework of the JNW naked singularity spacetime. Building upon prior research concerning black holes ( recent development in photon ring dimming around the black hole can be found in \cite{Nomura:2022zyy,Roy:2023rjk}), we seek to examine the influence of the absence of an event horizon on the likelihood of photon-axion conversion and its associated observational signals. We specifically compute the rate of photon-axion conversion in proximity to the photon ring and evaluate how this conversion results in a reduction of the  luminosity of the photon ring at high frequencies. The observed dimming, particularly evident in the X-ray and gamma-ray spectra, may serve as a potential observational indicator of axions near naked singularities.
}

{

The motivation behind this work is twofold. In the first place, we want to discern potential observational distinctions between black holes and naked singularities, utilizing the photon ring as a crucial instrument for probing the fundamental spacetime geometry. Secondly, we intend to establish a framework for identifying axion-like particles by their interaction with photons in intense magnetic fields, thereby aiding the extensive search for dark matter candidates. This amalgamation of objectives underscores the significance of this study in the realms of astrophysics and particle physics. {It is worthwhile to mention that the study of axionic particles around compact objects is an active area of research in contemporary physics\cite{Chen:2024nua,Chen:2022oad}.}

}

{

In summary, this paper provides a theoretical analysis of photon-axion conversion within the JNW naked singularity spacetime, emphasizing its observable implications for photon ring luminosity. Through the examination of the attenuation of the photon ring in the X-ray and gamma-ray spectra, we present a novel approach for differentiating naked singularities from black holes, while simultaneously offering a potential observational indicator of axion-like particles. With the advancement of observational tools, the proposed framework may facilitate novel discoveries in astrophysics and particle physics.

}

This paper is organised as follows: In \ref{SEC:PA_CONV}, comprehensive treatment of photon-axion conversion mechanism has been done.
  Concept of relative luminosity and conversion probability for the conversion mechanism have been studied in \ref{SEC:RLPA}. In the same section, we have also discussed the possibility of scattering of photons during the propagation. Next section , i.e \ref{SEC:JNW} is dedicated to the study of Janis-Newman-Winicour naked singularity and the conversion factor for such a spacetime. In second last section of paper, i.e, in \ref{SEC:DPRL}, we have explored the dimming phenomenon of the photon ring for JNW naked singularity followed by a discussion on required resolution to observe such dimming effect. Finally we have dedicated the last section  to the discussion of  the major findings and possible future research projects based on the photon-axion conversion phenomena.

\section{Photon-Axion conversion }\label{SEC:PA_CONV}
{ This section provides a concise overview of the photon-axion conversion phenomena in the presence of an external magnetic field\cite{Raffelt:1987im,Hochmuth:2007hk,Masaki:2017aea}. Conversion can occur efficiently for X-ray and gamma-ray radiation flowing near black holes, such as M87*. We take into account photons prpogating through an external magnetic field. Photons that have polarization parallel to the magnetic field are transformed into axions. Let us enumerate the parameters pertinent to the conversion:

\begin{enumerate}
	\item[(i)] Frequency of the photons progating through the magnetic field ($\omega$)
	\item[(ii)] Mass of the axion particle ($m_{\Phi}$)
	\item[(iii)] Photon-axion coupling constnat ($g_{\Phi\upgamma}$)
	\item[(iv)] Number density of the electron in the accreting medium ($n_{e}$)
	\item[(v)] Component of magnetic field perpendicular to the photon propagation ($|\mathbf{B}|$)
\end{enumerate}

The electron number density, $n_{e}$, is utilized to ascertain the plasma frequency.
}

  we can account for the influences of the surrounding plasma, with $\omega_{\rm pl}$ is plasma frequency,
\begin{equation}
    \omega_{\rm pl}\equiv\sqrt{\frac{4\pi\alpha n_e}{m_e}} = 3.7 \times 10^{-11}\text{eV}\sqrt{\frac{n_e}{\text{cm}^{-3}}},
\end{equation}
where $\alpha$ is the fine structure constant (with value $1/137$) and $m_e$ is the electron mass ($511$ keV).

{ Following the propagation of photons across a distance $z$, the probability of conversion from photons to axions is expressed as}

\begin{equation}
    \begin{aligned}
        P_{\upgamma\rightarrow\Phi}(z) 
        &= \left( \frac{\Delta_{\text{M}}}{\Delta_{\rm osc}/2} \right)^2 \sin^2{\Bigr( \frac{\Delta_{\rm osc}}{2}z \Bigl)},
    \end{aligned}\label{Eq:prob}
\end{equation}
where, $\Delta_{\rm osc}^{-1}$ represents the oscillation length, defined as
\begin{equation}\label{Eq:Def}
    \Delta_{\rm osc} = \sqrt{(\Delta_\Phi - \Delta_{\|})^2+(2\Delta_{\text{M}})^2}
\end{equation}
with
\begin{subequations}
	\begin{flalign}
		\Delta_{\|} &=\Delta_{\rm pl}-\Delta_{\rm vac},\\
		\Delta_{\rm pl} &=6.9 \times 10^{-25} \mathrm{eV}\left(\frac{n_e}{\mathrm{~cm}^{-3}}\right)\left(\frac{\mathrm{keV}}{\omega}\right),\\
		\Delta_{\rm vac} & = 9.3 \times 10^{-29} \mathrm{eV}\left(\frac{\omega}{\mathrm{keV}}\right)\left(\frac{|\mathbf{B}|}{\mathrm{Gauss}}\right)^2  ,\\
		\Delta_{\mathrm{M}} &= 9.8 \times 10^{-23} \mathrm{eV}\left(\frac{g_{\Phi \upgamma}}{10^{-11} \mathrm{GeV}^{-1}}\right)\Bigl(\frac{|\mathbf{B}|}{\text{ Gauss }}\Bigl) ,\\
		\Delta_{\Phi} & = 5 \times 10^{-22} \mathrm{eV}\left(\frac{m_{\Phi}}{\mathrm{neV}}\right)^2\left(\frac{\mathrm{keV}}{\omega}\right).
	\end{flalign}
\end{subequations}
These $\Delta_{\rm pl}$, $\Delta_{\rm vac}$, $\Delta_{\text{M}}$, $\Delta_{\Phi}$  have been evaluated in our analysis for M$87^*$ and are consistently defined as positive, which differs from the definition in \cite{Raffelt:1987im}.

 The conversion of photons to axions crucially depends on determining $\Delta_{\text{M}}$ using the parameters $g_{\Phi\upgamma}$ and $|\textbf{B}
 |$. The conversion process is hindered by the contributions of finite axion mass encoded in $\Delta_\Phi$, one-loop corrections of electrons encoded in $\Delta_{\rm vac}$, and plasma contribution encoded in $\Delta_{\rm pl}$. In the case of relativistic axions, the frequency of the propagating photons, i.e, $\omega$ is markedly greater than the axion mass $m_\Phi$, and for photons traversing through a medium, $\omega$ significantly exceeds the plasma frequency $\omega_{\rm pl}$. To confirm the validity of the current framework, a minimum of three of the following requirements must be fulfilled:
 \begin{enumerate}
 \renewcommand{\labelenumi}{\roman{enumi}.}
  \item 
\begin{flalign*}
  & \frac{\alpha}{45\pi}\left( \frac{|\mathbf{B}|}{|\mathbf{B}|_{\text{crit}}} \right)^2 \ll 1, \\
  &\text{where } |\mathbf{B}|_{\text{crit}} \equiv \frac{m_e^2}{\sqrt{4\pi\alpha}} = 4 \times 10^{13} \ \text{Gauss},
\end{flalign*}
where $m_{e}$ is the mass of the elctron and $\alpha$ is the structure constant.
  \item To validate the assumption of relativistic axions, it is necessary that
  \begin{equation*}
      \omega \gg m_\Phi,
  \end{equation*}
  \item For photons to move effectively in the surrounding plasma, it is essential that
  \begin{equation*}
      \omega \gg \omega_{\rm pl}. 
  \end{equation*}
\end{enumerate}

From \ref{Eq:prob} and \ref{Eq:Def}, one can notice that the condition for the most effective conversion probability is
\begin{flalign}
    \left(\Delta_{\Phi}+\Delta_{\rm vac}-\Delta_{\rm pl} \right)^{2} \ll 4\Delta_{M}^{2}
\end{flalign}
For such scenario, using \ref{Eq:Def}, one can find the characteristic length scale associated with the most effective conversion as
\begin{flalign}\label{Eq:EC}    \Delta_{\rm osc}^{-1}\simeq \left(2\Delta_{M} \right)^{-1}=3.4\times 10^{2}\times \left(2\times 10^{9}M_{\odot} \right)\times \left(\dfrac{\text{Gauss}}{|\mathbf{B}|} \right)\left( \dfrac{10^{-11}\rm Gev^{-1}}{g_{\Phi\upgamma}}\right)
\end{flalign}
{ The Event Horizon Telescope recently captured images of polarised synchrotron emission at 230 GHz from the compact object located at the centre of the M87 galaxy\cite{EventHorizonTelescope:2021bee}. EHT collaboration evaluated the magnetic field strength to be approximately $1-30$ Gauss, average electron density ($n_{e}$) around $10^{3}-10^{7}~\rm cm^{-3}$, and the electron temperature of the radiating plasma to be around $10^{11}\rm ~K$\cite{EventHorizonTelescope:2021srq}.} For supermassive compact object M$87^{*}$, the mass has been reported to be $6.2^{+1.1}_{-0.5}\times 10^{9} ~M_{\odot}$. Again from \ref{Eq:EC}, one can notice that for the magnetic field $|\mathbf{B}|\sim 10^{1-2}$ Gauss, the conversion length becomes comparable to Schwarzschild radius of a supermassive black hole of mass around $10^{9}~M_{\odot}$ if the  coupling constant is $g_{\Phi\upgamma}\sim 10^{-11}~ \rm GeV^{-1}$. So, for supermassive compact object M$87^{*}$, we can expect the conversion to axion to occur efficiently since photons can stay at the photon sphere for a considerable amount of time.

\section{Relative luminosity and conversion of photons into axions}\label{SEC:RLPA}

\subsection{Relationship between the spacetime metric and the conversion probability and conversion factor}

In the preceding section, we explored the presence of the photon sphere within a generally spherically symmetric spacetime and calculated the photon time-lapse in such contexts. To streamline our analysis, we adopt the universal spherically symmetric metric in the following form:
\begin{flalign}\label{Eq:SSM1}
ds^{2}=-f(r)dt^{2}+\dfrac{1}{f(r)}dr^{2}+r^{2}\mathcal{R}^{2}(r)\left(d\theta^{2}+\sin^{2}\theta d\phi^{2} \right)
\end{flalign}
The period during which photons remain near the photon sphere and their impact parameter $b$ is close to critical value  $b_{c}$ around the photon sphere ($r_{ph}<r<r_{ph}(1+\delta r)$) is given by \cite{Gralla:2019xty}
\begin{flalign}
    \mathfrak{T}(b)=-\dfrac{1}{f(r_{ph})}\dfrac{r_{ph}}{\sqrt{\mathcal{P}}}\ln\left[\dfrac{2(b-b_{c})}{\mathcal{P}b_{c}}\dfrac{r_{ph}^{2}}{\epsilon^{2}M^{2}} \right]
\end{flalign}

{ 
	where we have defined
	\begin{flalign}
		\mathcal{P}\equiv 1+r_{ph}^{2}\left[\left\{\dfrac{f'(r_{ph})}{f(r_{ph})}\right\}^{2}-3\left\{\dfrac{\mathcal{R}'(r_{ph})}{\mathcal{R}(r_{ph})}\right\}^{2} -\dfrac{2f'(r_{ph})}{r_{ph}f(r_{ph})}-\dfrac{f''(r_{ph})}{2f(r_{ph})}+\dfrac{\mathcal{R}''(r_{ph})}{\mathcal{R}(r_{ph})}\right] 
	\end{flalign}

and the small parameter $\delta r$ has been redefined as
\begin{flalign}
	\epsilon=\left(\dfrac{r_{ph}}{M} \right)\delta r
\end{flalign}
Here $M$ is the mass parameter of the given spacetime.

}

Now, to determine how many photons ($N_{\upgamma\rightarrow\Phi}$) near the photon sphere converted to axions per unit time $t$ and unit frequency $\omega_{c}$, we need to use the following equation\cite{Nomura:2022zyy,Roy:2023rjk}
\begin{flalign}
    \dfrac{\dd^{2}N_{\upgamma\to\Phi}}{\dd t\dd\omega_{c}}=\int_{b_{c}}^{b_{c}(1+\mathfrak{a}\epsilon^{2})} \dd b \dfrac{1}{2}\left(\dfrac{\dd^{3}N}{\dd t\dd\omega_{c}\dd b} \right) P_{\upgamma\to\Phi}\left(\sqrt{f(r_{ph})}\mathfrak{T}(b) \right)\label{Eq:3.6}
\end{flalign}

{
where
\begin{subequations}
\begin{flalign}
P_{\upgamma\to \Phi}\left(\sqrt{f(r_{ph})}\mathfrak{T} (b)\right)=\left(\dfrac{2\Delta_{M}}{\Delta_{\rm osc}} \right)^{2}\times \sin^{2}\left(-\dfrac{\Delta_{\rm osc}}{2}\dfrac{1}{\sqrt{f(r_{ph})}}\dfrac{r_{ph}}{\sqrt{\mathcal{P}}}\ln\left[\dfrac{2(b-b_{c})}{\mathcal{P}b_{c}}\dfrac{r_{ph}^{2}}{\epsilon^{2}M^{2}} \right] \right)\\
	\bigg[\dfrac{1}{8}\left\{\dfrac{4r_{ph}f'(r_{ph})}{f(r_{ph})}+\left(\dfrac{r_{ph}f'(r_{ph})}{f(r_{ph})}\right)^{2}-2r_{ph}^{2}\left(\dfrac{f''(r_{ph})}{f(r_{ph})}-\dfrac{2\mathcal{R}''(r_{ph})}{\mathcal{R}(r_{ph})}\right)\right\}-1\bigg]\left(\dfrac{M}{r_{ph}} \right)^{2}\equiv \mathfrak{a}
\end{flalign}
\end{subequations}}

The factor of $1/2$ in \ref{Eq:3.6} represents the proportionality between photons with polarisation parallel to the external magnetic field and the production of axions. The number of photons entering the region $r_{ph}<r<r_{ph}\left(1+\epsilon M/r_{ph} \right)$ and escaping to infinity can be found by integrating over the interval $(b_{c},,b_{c}(1+\mathfrak{a}\epsilon^{2}))$. Since there is no significant variation in the number of photons when the impact parameter changes in this situation, we can approximate it by considering its value at $b=b_{c}$ outside the integral. Then \ref{Eq:3.6} takes the following form

\begin{flalign}
   \dfrac{\dd^{2}N_{\upgamma\to \Phi}}{\dd t \dd \omega_{c}}\simeq \dfrac{1}{2}\left(\dfrac{\dd^{3}N}{\dd t\dd \omega_{c}\dd b} \right)\bigg\vert_{b}\times \left(\dfrac{2\Delta_{\rm M}}{\Delta_{\rm osc}} \right)^{2}\int_{b_{c}}^{b_{c}(1+\mathfrak{a}\epsilon^{2})}\sin^{2}\left[\dfrac{\Delta_{\rm osc}}{2}\mathfrak{T}(b)\sqrt{f(r_{ph})} \right]\dd b \label{Eq:3.8}
\end{flalign}

The integral in \ref{Eq:3.8} can further be simplified as
\begin{flalign}
    \mathcal{I}=\int_{b_{c}}^{b_{c}(1+\mathfrak{a}\epsilon^{2})}\dd b ~\sin^{2}\left[\dfrac{\Delta_{\rm osc}}{2}\mathfrak{T}(b)\sqrt{f(r_{ph})} \right]=\dfrac{1}{2}\left(\dfrac{1}{1+\mathcal{Y}^{2}} \right)\mathfrak{a}b_{c}\epsilon^{2}
\end{flalign}
where we have used the following definition
\begin{flalign}
    \mathcal{Y}=\dfrac{\sqrt{\mathcal{P}}\sqrt{f(r_{ph})}}{r_{ph}\Delta_{\rm osc}}
\end{flalign}

The fraction of photons that transform into axions when entering the vicinity of the photon sphere is provided as 
\begin{flalign}\label{conversion factor}
		\frac{\dd^{2}N_{\upgamma \rightarrow \phi}}{\dd t \dd\omega_c}\bigg/\dfrac{\dd^{2}N}{\dd t\dd\omega_{c}}\simeq \dfrac{1}{4} \left(\frac{2\Delta_{\rm M}}{\Delta_{\rm osc}}\right)^{2}  \left(\dfrac{1}{1+\mathcal{Y}^{2}} \right) 
\end{flalign}

Now, in the scenario of efficient conversion, we can choose the condition to be $\Delta_{\rm osc}\simeq 2\Delta_{\rm M}$. In this particular case, the conversion factor(C.F.) can be given by
\begin{flalign}
		CF=	\frac{\dd N_{\upgamma \rightarrow \Phi}}{\dd t \dd\omega_c}\bigg/\dfrac{\dd^{2}N}{\dd t\dd\omega_{c}} \Bigg|_{\text{ efficient conversion}}&\simeq \dfrac{1}{4}\left(\dfrac{1}{1+\mathcal{Y}^{2}} \right)\\
			&=\dfrac{1}{4}\left[\dfrac{\left(\dfrac{r_{ph}}{M} \right)^{2} \left(2 M\Delta_{\rm M}\right)^{2}}{\left(\dfrac{r_{ph}}{M} \right)^{2} \left(2 M\Delta_{\rm M}\right)^{2}+{\mathcal{P} f(r_{ph})}} \right]\label{Eq:CF_end} 
\end{flalign}

 Now, inserting \ref{eq:4.9} into the \ref{Eq:3.8}, we have
\begin{flalign}\label{Eq:Axion_Num1}	\dfrac{\dd^{2}N_{\upgamma\to\Phi}}{\dd t\dd\omega_{c}}\simeq &{\mathfrak{a} \pi^{2}\epsilon^{2}}b_{c}\left(\dfrac{2\Delta_{\rm M}}{\Delta_{\rm osc}} \right)^{2}\left[\dfrac{\left(\Delta_{\rm osc}r_{ph}\right)^{2}}{\left(\Delta_{\rm osc}r_{ph}\right)^{2}+\mathcal{P} f(r_{ph})} \right] \nonumber\\
&\times \int_{r_{\rm in}} ^{r_{\rm out}} \dd r_e  J_e\left(\frac{\sqrt{f(r_{ph})}\omega_c}{\sqrt{f(r_e)}},r_e \right)\left[ \frac{br_e\mathcal{R}(r_{e})\sqrt{f(r_e)}}{\sqrt{\dfrac{r_e^2\mathcal{R}^{2}(r_{e})}{f(r_e)}-b^2}} \right]
\end{flalign}
\ref{Eq:Axion_Num1} helps us to determine the number of axions produced during the conversion of photons into axion.

\subsection{Scattering of photons by the plasma}\label{SEC:TSS}
In earlier analysis, for the shake of simplicity,  we have neglected the possibility of the scattering of photons by the surrounding plasma while it propagate through the plasma.
While taking into consideration the fact that photons have a limited mean free path while incorporating the scattering phenomena, the expression for the number of photons that are converted into axions per unit time and unit frequency as given in \ref{Eq:3.6} needs to be modified in the following way
\begin{flalign}
\dfrac{\dd^{2}N_{\upgamma\to\Phi}}{\dd t \dd \omega}=\int_{b_{c}}^{b_{c}(1+\mathfrak{a}\epsilon^{2})} \dd b \dfrac{1}{2}\left(\dfrac{\dd^{3}}{\dd t \dd \omega_{c}\dd b} \right)P_{\upgamma\to \Phi}\left(\sqrt{f(r_{ph})} \mathfrak{T}(b)\right) \exp\left(-\dfrac{\sqrt{f(r_{ph})}\mathfrak{T}(b)}{\ell} \right)
\end{flalign}
where $\ell$ denotes the mean free path of the propagating photons. Implementing similar treatment like \ref{Eq:3.8}, one can come up with the following equation
\begin{flalign}
\dfrac{\dd^{2}N_{\upgamma\to\Phi}}{\dd t \dd \omega}=\dfrac{1}{2}\left(\dfrac{\dd^{3}N}{\dd t \dd \omega_{c}\dd t} \right)\bigg\vert_{b=b_{c}}\times \left(\dfrac{2\Delta_{M}}{\Delta_{\rm osc}} \right)^{2}\times \int_{b_{c}}^{b_{c}(1+\mathfrak{a}\epsilon^{2})}\dd b \sin^{2}\left[\dfrac{\Delta_{\rm osc}}{2}\mathfrak{T}(b)\sqrt{f(r_{ph})} \right] \exp\left(-\dfrac{\sqrt{f(r_{ph})}\mathfrak{T}(b)}{\ell} \right)\label{Eq:INF1}
\end{flalign}
Let us define the quantity
\begin{flalign}
\mathcal{Z}\equiv \dfrac{1}{\ell \Delta_{\rm osc}}
\end{flalign}
Hence the number of photons converted into axions per unit time per unit frequency is provided by
\begin{flalign}
\dfrac{\dd^{2}N_{\upgamma\to \Phi}}{\dd t \dd \omega}&=\dfrac{\mathfrak{a}b_{c}\epsilon^{2}}{4}\left(\dfrac{\dd^{3}N}{\dd t \dd\omega_{c}\dd t} \right)\bigg\vert_{b=b_{c}}\left(\dfrac{2\Delta_{M}}{\Delta_{\rm osc}} \right)^{2}\left[\left(1+\dfrac{\mathcal{Z}}{\mathcal{Y}} \right)\left\{1+(\mathcal{Y+Z})^{2} \right\} \right]^{-1}\\
&=\dfrac{\mathfrak{a}b_{c}\epsilon^{2}}{4}\left(\dfrac{\dd^{3}N}{\dd t \dd \omega_{c}\dd t} \right)\bigg\vert_{b=b_{c}}\left(\dfrac{2\Delta_{M}}{\Delta_{\rm osc}} \right)^{2}\left[\left(1+\dfrac{1}{\left(\dfrac{\ell}{r_{ph}}\right) \sqrt{\mathcal{P}}\sqrt{f(r_{ph})}} \right)\left\{1+\left(\dfrac{1+\left(\dfrac{\ell}{r_{ph}}\right) \sqrt{\mathcal{P}}\sqrt{f(r_{ph})}}{\left(\dfrac{ \ell}{r_{ph}}\right) ~ \left(\dfrac{r_{ph}}{M}\right)M\Delta_{\rm osc}}\right)^{2} \right\} \right]^{-1}\label{Eq:CF_Plasma}
\end{flalign}

It is important to note that we can retrieve \ref{Eq:CF_end} with help of \ref{Eq:CF_Plasma} when the following condition holds
\begin{flalign}
\dfrac{\ell/M}{\dfrac{1}{\sqrt{\mathcal{P}}\sqrt{f(r_{ph})}}\left(\dfrac{r_{ph}}{M}\right)} \ggg 1 \label{Eq:TS}
\end{flalign}
The mean free path for photons with a frequency lower than the electron mass $m_{e}$ is provided by
\begin{flalign}
    \dfrac{\ell}{M}=\dfrac{1}{\left(\dfrac{\sigma_{T}}{M^{2}}\right)\left(n_{e}M^{3}\right)}
\end{flalign}
where $\sigma_{T}=\dfrac{8\pi}{3}\left(\dfrac{\alpha}{m_{e}} \right)^{2}\approx 1.3\times 10^{-25}~ M_{\odot}^{2}$ is the \emph{Thomson scattering cross-section} and $n_{e}$ is the electron density. In our analysis, we have considered the compact object which has billions of solar mass. For such scenario, it is clear that \ref{Eq:TS} is automatically  satisfied if 
\begin{flalign}
  \mathfrak{S}\equiv  \dfrac{1}{\sqrt{\mathcal{P}}\sqrt{f(r_{ph})}}\dfrac{r_{ph}}{M}\lesssim 10\label{Eq:RQ}
\end{flalign}
From now we denote this important factor $\mathfrak{S}\equiv\dfrac{1}{\sqrt{\mathcal{P}}\sqrt{f(r_{ph})}}\dfrac{r_{ph}}{M}$ as \emph{"scattering-limit-factor"}. In the next section, we will show that \ref{Eq:RQ} is indeed satisfied in our analysis.

\section{Janis-Newman-Winicour Spacetime: An overview in brief}\label{SEC:JNW}

{ In the present work we discuss the Janis-Newman-Winicour (JNW) naked singularity which represents an exact solution of the Einstein’s equations with a massless scalar field\cite{PhysRevLett.20.878}. Fisher developed this solution initially using a different parametrisation\cite{Fisher:1948yn}, and Bronnikov and  Khodunov later investigated its stability\cite{10.1093/mnras/215.4.575}. Virbhadra proved the equivalency of the Wyman solution with the Janis-Newman-Winicour spacetime\cite{Virbhadra:1997ie}, which Wyman discovered  earlier\cite{PhysRevD.24.839}. Interestingly, the spherically symmetric and asymptotically flat exact metric solution becomes a naked singularity solution rather than a Schwarzschild solution upon the inclusion of the massless scalar field to the action. The optical characteristics of the Janis-Newman-Winicour spacetime, such as gravitational lensing, accretion, and shadow, have been the subject of numerous literary works\cite{Gyulchev:2008ff,PhysRevD.65.103004,Virbhadra:1998dy,Virbhadra:2007kw,Yang:2015hwf,Takahashi:2004xh,PhysRevD.100.024055,Chowdhury:2011aa,Sau:2020xau,Chen:2023uuy,Stashko:2022dtx,Patel:2022vlu}.  }

\subsection{Revisiting the spacetime}
In this study, we explore the Einstein massless scalar field theory (EMS) with minimal coupling between the massless scalar field and gravity.  The action associated to the theory is expressed by\cite{PhysRevD.24.839,Virbhadra:1997ie}
\begin{flalign}
    S=\int \dd^{4}x \sqrt{-g} \left[\dfrac{R}{16\pi G} -\dfrac{1}{2}\nabla_{\mu}\xi(r)\nabla^{\mu}\xi (r)\right]
\end{flalign}
where, $g$ is the determinant of the metric tensor, $R$ is the Ricci scalar and $\xi(r)$ is the minimally coupled non-trivial scalar field. The associated Einstein gravitational field equations, which are derived from the action above, have an exact static solution that is spherically symmetric in four dimensions. The corresponding spacetime metric is given byx
\begin{flalign}
    ds^{2}=-\left(1-\dfrac{B}{r} \right)^{\gamma}dt^{2}+\left(1-\dfrac{B}{r} \right)^{-\gamma} dr^{2}+\left(1-\dfrac{B}{r} \right)^{1-\gamma} r^{2}\left(d\theta^{2}+\sin\theta d\phi^{2} \right)
\end{flalign}
where the range of the parameter $\gamma$ is $0\leq \gamma <1$ and $B=\dfrac{2M}{\gamma}$, $M$ being the  Arnowitt-Deser-Misner (ADM) mass of the gravitating object. One can notice that when $\gamma=1$, the solution becomes a Schwarzschild solution.
This spacetime is widely known as {\textbf{Janis-Newman-Winicour}} solution. At $r=B$, there exists a curvature singularity. We limit ourselves to the region $r>B$ since this metric depicts a naked singularity because it is not covered by the event horizon for a non-trivial scalar field. The scalar field solution and the corresponding energy-momentum tensor are provided, respectively, by\cite{Sau:2020xau}
\begin{flalign}
    \xi(r)&=\dfrac{Q}{B}\ln\left(1-\dfrac{B}{r} \right)\\
    T_{\mu\nu}&=\nabla_{\mu}\xi\nabla_{\nu}\xi
-\dfrac{1}{2}g_{\mu\nu}\nabla^{\alpha}\xi\nabla_{\alpha}\xi
\end{flalign}
where parameter $B$ is related to the scalar charge $Q$ by
\begin{flalign}
    B=2\sqrt{Q^{2}+M^{2}}=2M\sqrt{1+q^{2}}
\end{flalign}
where $q\equiv \dfrac{Q}{M}$ is the dimensionless charge parameter. It is easy to notice that with the vanishing charge parameter, one can recover the Schwarzschild solution. For the existence of a photon sphere, the range of the charge parameter is $0<q\leq\sqrt{3}$, which is equivalent to the parameter space $0.5\leq \gamma \leq 1.0$. The existence of the photon sphere and the corresponding shadow cast will be discussed in the following subsection.

\subsection{Shadow cast by Janis-Newman-Winicour naked singularity}
The literature on the shadow of naked singularity is extensive\cite{Vagnozzi:2022moj,Sau:2020xau,Solanki:2021mkt}, much like that on black hole shadows. Indeed, research has been done on the shadow of naked singularity even in the absence of a photon sphere\cite{Joshi:2020tlq}. Nevertheless, we will limit our region of analysis to, $0.5<\gamma\leq1$ for obvious reasons. In this particular limit, we will have both the photon sphere and shadow cast for spacetime. However, one can extend the limit of the parameter $\gamma$ to study the other properties of spacetime. 

\begin{figure}[ht]
    \centering 
    \includegraphics[scale=0.80]{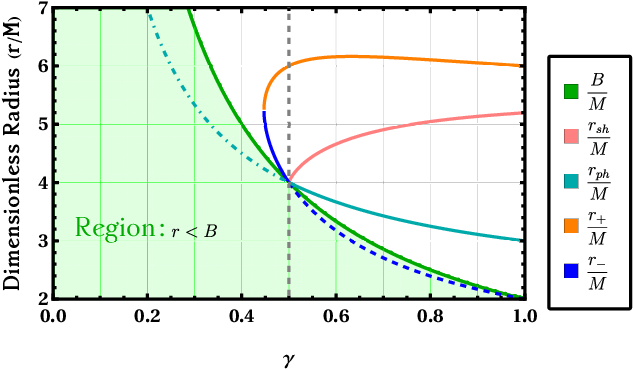}
    \caption{\textit{The figure above shows how the radius of the photon sphere ($r_{ph}$), the corresponding shadow radius ($r_{sh}$), and the innermost circular orbits ($r_{\pm}$) for massive particles in \textbf{Janis-Newman-Winicour (JNW)} spacetime depend on the parameter $\gamma$. Unlike other spacetimes, in this scenario, the shadow radius decreases monotonically as the photon sphere radius increases within the range $0.5\leq\gamma\leq 1$. The JNW spacetime features a singularity at $r=B\equiv r_{g}$, where $B=\dfrac{2M}{\gamma}$, $M$ being the Arnowitt-Deser-Misner (ADM) mass of the gravitating object. The green shaded area, i.e., the region $r<r_{g}=B$, represents an unphysical solution. Additionally, at $\gamma = 0.50$, the photon sphere radius and shadow radius coincide, meaning $r_{ph}=r_{sh}=r_{g}=B$. The dashed blue line and dot-dashed cyan lines depict extended numerical solutions of $r_{-}$ and $r_{ph}$, which are unphysical.}}
    \label{Fig:Radius}
\end{figure}

 The radius of photon sphere  $r_{ph}$ and the radius of the shadow $r_{sh}$ can be expressed in terms of the characteristic parameter $\gamma$ only. These are given as
\begin{subequations}
\begin{flalign}
    \dfrac{r_{ph}}{M}&= \left(2+\dfrac{1}{\gamma} \right)\\
    \dfrac{r_{sh}}{M}&=\sqrt{\left(2+\dfrac{1}{\gamma} \right)^{1+2\gamma}\left(2-\dfrac{1}{\gamma} \right)^{1-2\gamma}}
\end{flalign}
\end{subequations}
In \ref{Fig:Radius}, the atypical behaviour of the photon sphere and the corresponding shadow have been shown. One can notice that unlike most of the other scenarios, here for JNW spacetime, the radius of photon sphere increases and the shadow radius decreases as we gradually decrease the value of the characteristic parameter $\gamma$ in the range $\gamma \in [0.5,1)$. This counter-intuitive nature of photon sphere and shadow for JNW spacetime has been discussed widely in literature\cite{Sau:2020xau}.

We also need to discuss the innermost circular orbits for timelike geodesics in order to incorporate the behaviour of the accretion disk with the change of the characteristic parameter $\gamma$. Depending on the range of the parameter, there can be either one or two circular timelike geodesics in the equatorial plane. These are calculated as the roots of the following equation:
\begin{flalign}
    \left(\dfrac{r}{M}\right)^{2}-2\left(3+\dfrac{1}{\gamma} \right)\dfrac{r}{M}+2\left(2+\dfrac{3}{\gamma}+\dfrac{1}{\gamma^{2}} \right)=0\label{Eq:Roots}
\end{flalign}
Solutions of the \ref{Eq:Roots} can  explicitly  be found as
\begin{flalign}
    \dfrac{r_{\pm}}{M}=3+\dfrac{1}{\gamma}\pm \sqrt{5-\dfrac{1}{\gamma^{2}}}\label{Eq:Roots_Sol}
\end{flalign}
In \ref{Fig:Radius}, the radius of the outer circular orbit ($r_{+}$) has been shown with an orange solid line, and the radius of the inner circular orbit ($r_{-}$) has been depicted with a blue solid line. From the figure, it is easy to understand that for the range $\gamma\in [0.5,1)$, only one circular orbit exists. The radius of this innermost circular orbit initially increases with a decreasing value of $\gamma$ from unity, before getting decreased; however, the radius finally becomes the same for the Schwarzschild scenario for the value of $\gamma=0.5$. 
\subsection{Conversion factor for JNW spacetime}
\begin{figure}[htbp]
    \centering
   \subfloat[\emph{Change in the scattering-limit factor has been shown for the range of the characteristic parameter $\gamma$ in range $(0.5,1)$.} ]{{\includegraphics[width=7.3cm]{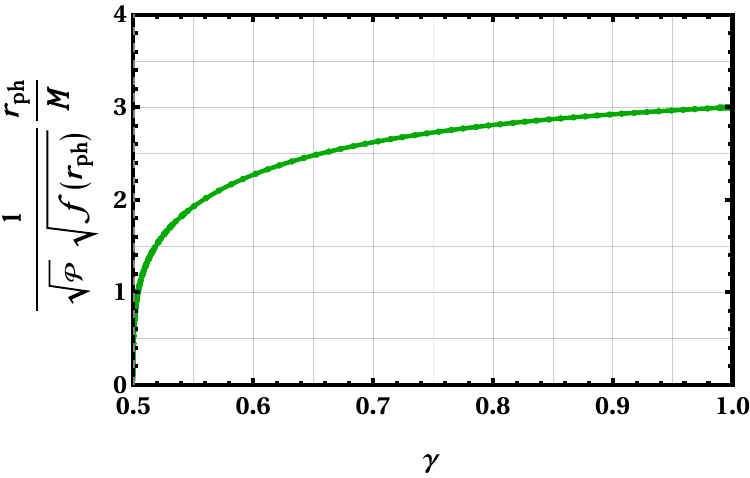}}\label{Fig:TSS_1}}
   \qquad
   \subfloat[\emph{Change in the scattering-limit factor has been shown for the range of the characteristic parameter $\gamma$ in range $(0.5,0.510)$.}]{{\includegraphics[width=7.8cm]{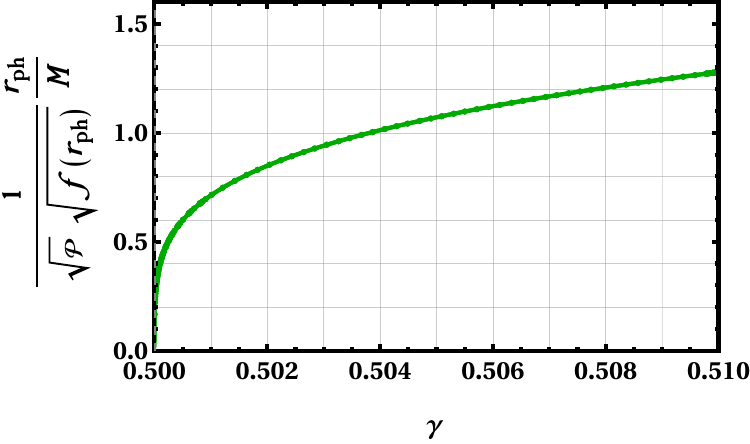}}\label{Fig:TSS_2}}
    \caption{\emph{Variation of the dimensionless scattering-limit-factor has been depicted with the change of the characteristic parameter $\gamma$ in the range $0.5\leq \gamma \leq 1$. One can observe that this factor monotonically increases from $0$ to $3$.}}
    \label{Fig:TSS_Variation}
\end{figure}

As discussed in previous section, the scattering of photon in plasma can affect the conversion factor. However, in \ref{SEC:TSS}, we also have discussed the possibility of neglecting the effect of such phenomena.  For JNW spacetime in the parameter range $0.5\leq \gamma \leq 1$, one can always show that
\begin{flalign}
    0\leq \dfrac{1}{\sqrt{\mathcal{P}}\sqrt{f(r_{ph})}}\dfrac{r_{ph}}{M}=\left(2-\dfrac{1}{\gamma} \right)^{\frac{1-\gamma}{2}}\left(2+\dfrac{1}{\gamma} \right)^{\frac{1+\gamma}{2}}\leq 3
\end{flalign}
This result has also been depicted in \ref{Fig:TSS_Variation}. As a result, for this spacetime, one can readily show that \ref{Eq:RQ} is indeed satisfied. Hence, we can ignore the scattering of photons inside the plasma for our further analysis.

\begin{figure}[htbp]
    \centering
   \subfloat[\emph{Rapid increase in conversion factor has been depicted for minor variation of the characteristic parameter $\gamma$ near $\gamma=0.50$ for different fixed values of dimensionless quantity $M\Delta_{M}$.}]{{\includegraphics[width=4.5cm]{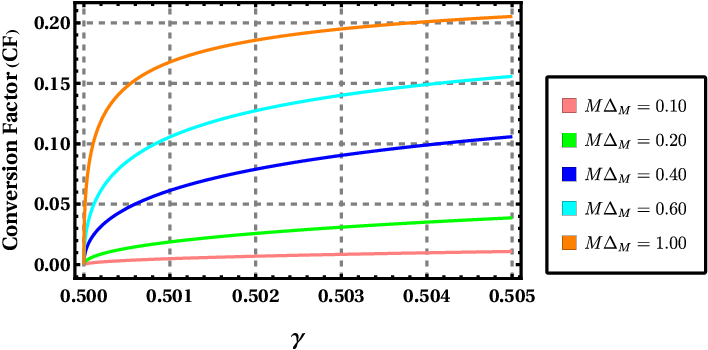}}\label{Fig:CF_a}}
   \qquad
   \subfloat[\emph{Variation of the conversion factor with the change in characteristic parameter $\gamma$ has been shown for full range of the parameter, for various values of dimensionless quantity $M\Delta_{M}$.}]{{\includegraphics[width=4.5cm]{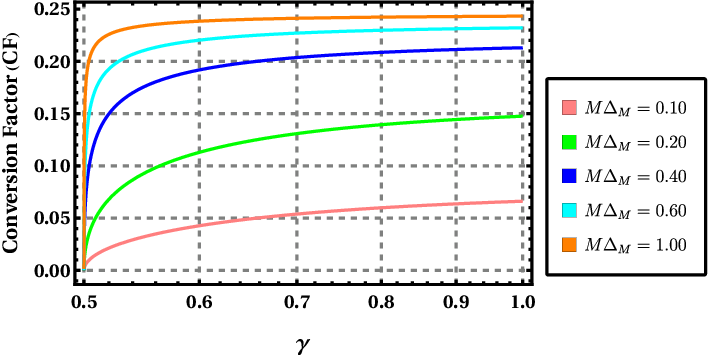}}\label{Fig:CF_b}}
   \qquad
   \subfloat[\emph{With changes in the dimensionless quantity $M\Delta_{M}$, variations in the conversion factor have been demonstrated for different values of dimensionless parameter $\gamma$.}]{{\includegraphics[width=4.5cm]{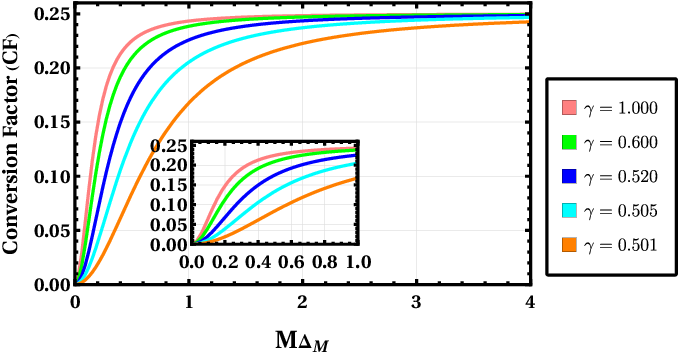}}\label{Fig:CF_c}}
\qquad\\
   \subfloat[\emph{This plot depicts the variation of the conversion factor in response to changes in two dimensionless parameters, $M\Delta_{M}$ and $\gamma$. Contour lines indicate constant levels of the conversion factor, highlighting the regions of the parameter space where significant changes in the conversion factor occur.}]{{\includegraphics[width=4.5cm]{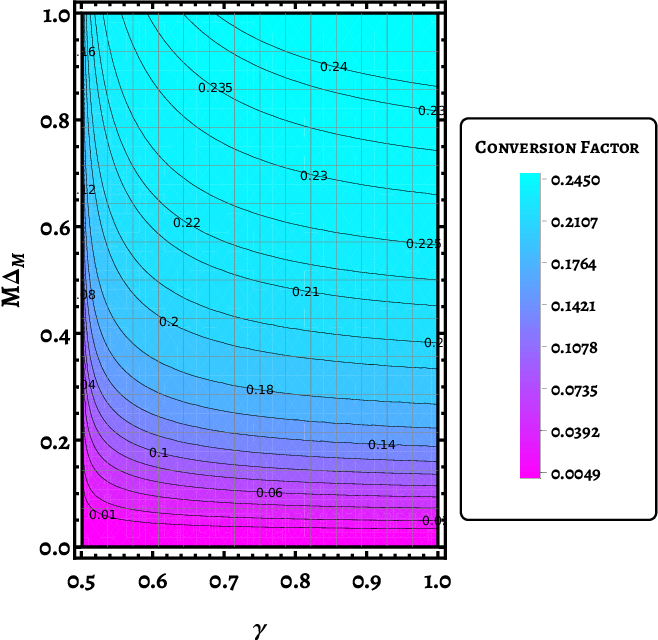}}\label{Fig:CF_d}}
   \qquad
   \subfloat[\emph{This density plot illustrates the distribution of the conversion factor over the ranges of two dimensionless parameters, $M\Delta_{M}$ and $\gamma$. The colour gradient represents the density of the conversion factor,  providing insights into the regions with significant variation in conversion efficiency.}]{{\includegraphics[width=4.5cm]{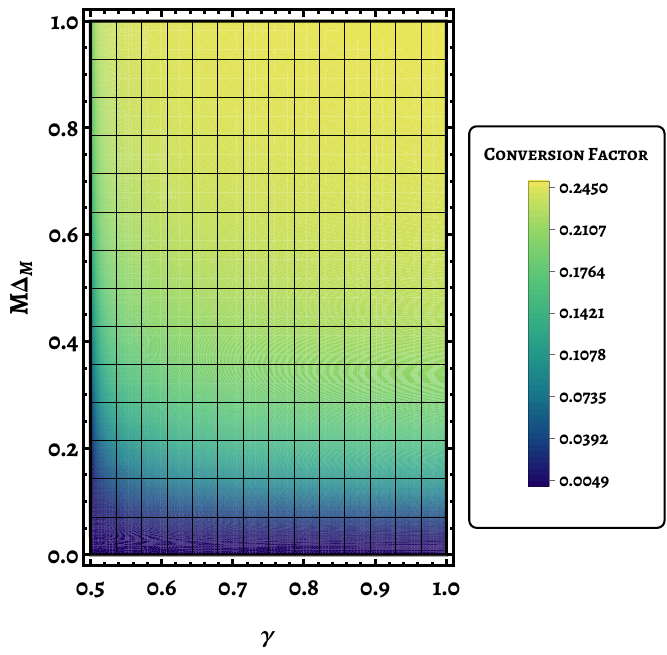}}\label{Fig:CF_e}}
    \caption{In the instance of effective conversion, i.e. when $\Delta_{\rm osc}=2\Delta_{M}$, the variation of the conversion factor has been demonstrated in conjunction with the change of the dimension-less parameters $\gamma$ and dimensionless quantity $M\Delta_{M}$.}
    \label{Fig:CF_Variation}
\end{figure}

In \ref{Fig:CF_Variation}, we have illustrated the variations of the conversion factor (CF), in the effective conversion regime, i.e, when $\Delta_{\rm osc}=2\Delta_{M}$, with the variation of dimensionless quantity $M\Delta_{\rm M}$ and dimensionless characteristic parameter $\gamma$. 

In particular,  in \ref{Fig:CF_a} and \ref{Fig:CF_b}, the variation of the conversion factor has been demonstrated while the characteristic parameter $\gamma$ has been varied, keeping the dimensionless quantity $M\Delta_{M}$ constant. One can observe that the change in conversion factor is more rapid in the range $0.50\lesssim \gamma \lesssim 0.60$. Although, for $\gamma>0.60$, the conversion factor monotonically increases, the rate of increase is much lower than  the previously mentioned range of the parameter $\gamma$. As shown in \ref{Fig:CF_c}, for a fixed value of the characteristic parameter $\gamma$, the conversion factor increases monotonically with the increase in the value of dimensionless quantity $M\Delta_{M}$, and gradually approaches its peak value i.e. $25\%$ conversion. This phenomenon is explained by the selective conversion of photons that have a polarization aligned with the magnetic field. The production of photons with this polarization and axions occurs at an equal rate due to significant mixing. Examining \ref{Fig:CF_c}, it can also be noticed that for sufficiently high value of dimensionless quantity $M\Delta_{M}$ (in this case $M\Delta_{M} \sim 4-5$), the conversion rate almost approaches its peak value, seemingly irrespective of the value of the characteristic parameter $\gamma$. This can also be confirmed by examining \ref{Eq:CF_end}. 

In \ref{Fig:CF_d}, contour lines represent consistent levels of the conversion factor, emphasizing the areas in the parameter space where notable fluctuations in the conversion factor take place. Similarly, in \ref{Fig:CF_e}, the colour gradient illustrates the density of the conversion factor, providing insights into the places with large variance in conversion efficiency. From \ref{Fig:CF_d} and \ref{Fig:CF_e}, it is easily understood that the  conversion factor is effectively  enhanced for relatively high value of both $\gamma$ and $M\Delta_{M}$.

\section{Dimming of the photon ring Luminosity}\label{SEC:DPRL}

\subsection{Dimming of the photon sphere}
After we have determined the conversion factor, all that is required of us is to make a count of the photon numbers that have approached the photon sphere and then escaped to infinity. This will allow us to determine the number of photons that have been converted and the newly formed axion. With the help of \ref{eq:4.9} and \ref{eq:4.11}, the photon count can be determined as
\begin{flalign}
    \dfrac{\dd^{2}N}{\dd t \dd\omega_{c}}=\mathfrak{L}^{0}_{\omega}\omega_{\rm obs}^{-1}\times 2\mathfrak{a}\left( \dfrac{b_{c}}{M}\right)^{2}\left(\dfrac{r_{ph}}{M} \right)\times \int_{x_{\rm in}}^{x_{\rm out}}\dd x_{e}\left[ \dfrac{x_{e}^{-3/2}\mathcal{R}(x_{e})f(x_{e})}{\sqrt{\dfrac{x_{e}^{2}\mathcal{R}^{2}(x_{e})}{f(x_{e})}-\left( \dfrac{b_{c}}{r_{ph}}\right)^{2}}}\right]e^{-\dfrac{\omega_{\rm obs}}{T_{e,c}}\dfrac{x_{e}}{\sqrt{f(x_{e})}}}\label{Eq:Suppression}
\end{flalign}
where the variable $x$ is defined as $x=r/r_{ph}$ and the quantity $\mathfrak{L}^{0}_{\omega}$ is defined as
\begin{flalign}
\mathfrak{L}^{0}_{\omega}&=2\pi^{2}\epsilon^{2}M^{3}\left(\dfrac{4\alpha^{3}}{3\pi m_{e}} \right)\left(\dfrac{2\pi}{3m_{e}} \right)^{1/2}T_{e,c}^{-1/2}n_{e,c}^{2}\bar{g}_{ff}\\
&=6.48\times 10^{27} {\rm erg.sec^{-1}.KeV^{-1}}\epsilon^{2}\left(\dfrac{M}{6.2\times 10^{9} M_{\odot}} \right)^{3}\left( \dfrac{T_{e,c}}{10^{11}K}\right)^{-1/2}\left(\dfrac{n_{e,c}}{10^{4}{\rm cm^{-3}}} \right)^{2}\bar{g}_{ff}
\end{flalign}
One should notice the fact, due to gravitational redshift, the observed frequency $\omega_{\rm obs}$ is directly connected to the frequency $\omega_{c}$ at the photon sphere as $\omega_{\rm obs}=\sqrt{f(r_{ph})}\omega_{c}$. As the distance of M87* is around $D\approx 16.8 {\rm Mpc}$, the cosmological redshift of M87* can be calculated to be $\mathfrak{z}\approx \dfrac{H_{0}D}{c}\approx 0.004$, where $H_{0}$ is the present value of the Hubble parameter, which is considered to be $71 {\rm km/s/Mpc}$ for our study. Within the local universe, galaxies, such as the Messier galaxy, generally have peculiar velocities that are normally in the range of a few hundred kilometres per second. However, while evaluating the prevailing gravitational redshift, we disregard minor factors like peculiar velocities and cosmic expansion.

We construct the following quantity to be the relative luminosity (RL) of photons prior to any conversion as
\begin{flalign}
    \dfrac{\mathfrak{L}^{\upgamma}_{\omega}}{\mathfrak{L}^{0}_{\omega}}\equiv\dfrac{\omega_{\rm obs}}{\mathfrak{L}^{0}_{\omega}}\left(\dfrac{\dd^{2}N}{\dd t\dd \omega_{c}} \right)=2\mathfrak{a}\left(\dfrac{b_{c}}{M} \right)^{2}\left(\dfrac{r_{ph}}{M} \right)\times\int_{x_{\rm in}}^{x_{\rm out}}\dd x_{e}\left[ \dfrac{x_{e}^{-3/2}\mathcal{R}(x_{e})f(x_{e})}{\sqrt{\dfrac{x_{e}^{2}\mathcal{R}^{2}(x_{e})}{f(x_{e})}-\left( \dfrac{b_{c}}{r_{ph}}\right)^{2}}}\right]e^{-\dfrac{\omega_{\rm obs}}{T_{e,c}}\dfrac{x_{e}}{\sqrt{f(x_{e})}}}
\end{flalign}
The relative luminosity for the axions as generated throughout photon-axion conversion can be calculated by
\begin{flalign}
    \dfrac{\mathfrak{L}^{\upgamma\to\Phi}_{\omega}}{\mathfrak{L}^{0}_{\omega}}=\dfrac{\mathfrak{L}^{\upgamma}_{\omega}}{\mathfrak{L}^{0}_{\omega}}\dfrac{1}{4}\left(\dfrac{2\Delta_{\rm M}}{\Delta_{\rm osc}} \right)^{2}\left(\dfrac{1}{1+\mathcal{Y}^{2}} \right)
\end{flalign}

\begin{figure}[htbp]
    \centering
   \subfloat[\emph{The relative brightness spectra of photons, both dimmed and undimmed, have been shown with the axion mass assumed to be $m_{\Phi}=1~~ \rm neV$ for different set of values of characteristic parameter $\gamma$.}]{{\includegraphics[width=7cm]{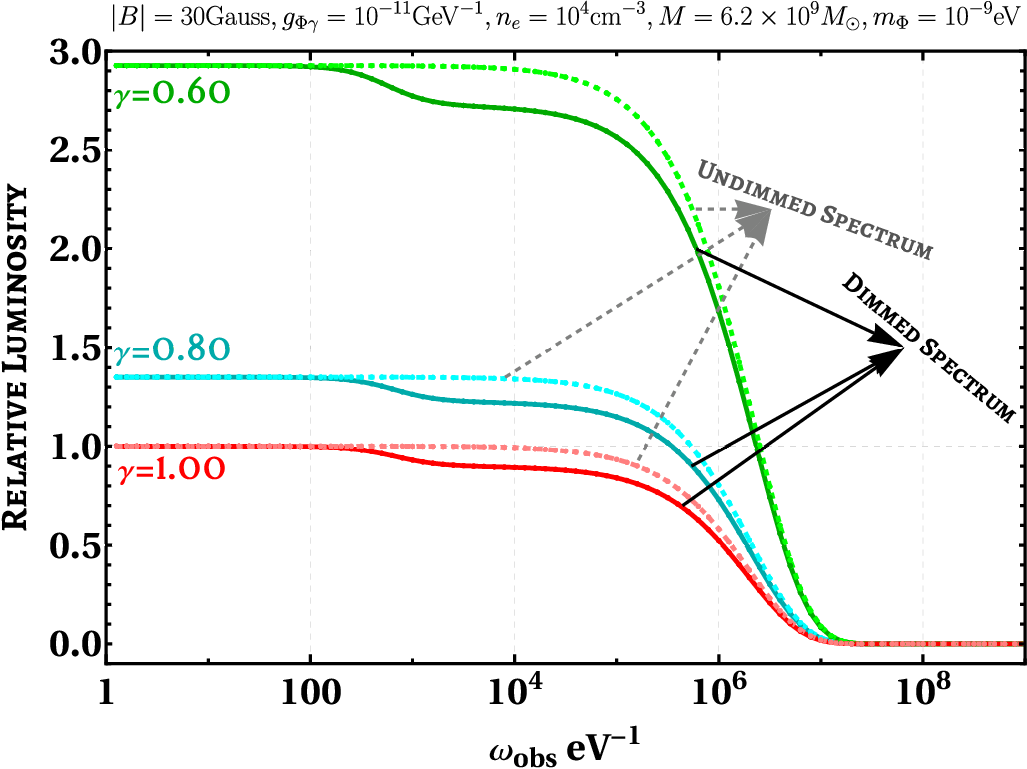}}\label{RLPH_a}}
   \qquad
   \subfloat[\emph{The relative luminosity spectrum of photons has been illustrated, considering the axion mass as $m_{\Phi}=10~~\rm neV$, with both dimmed and undimmed conditions for different set of values of characteristic parameter $\gamma$.}]{{\includegraphics[width=7cm]{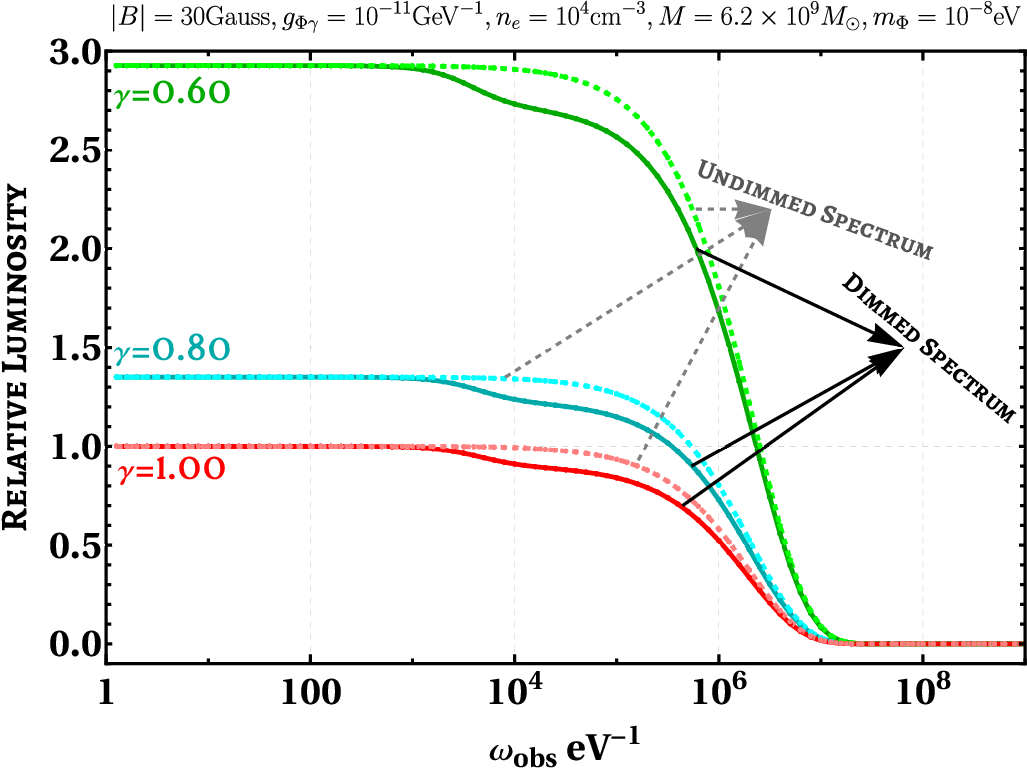}}\label{RLPH_b}}
   \qquad
   \subfloat[\emph{The relative brightness spectrum of axions is shown with an axion mass of $m_{\Phi}=1~~\rm neV$, with varying values of the characteristic parameter $\gamma$.}]{{\includegraphics[width=7cm]{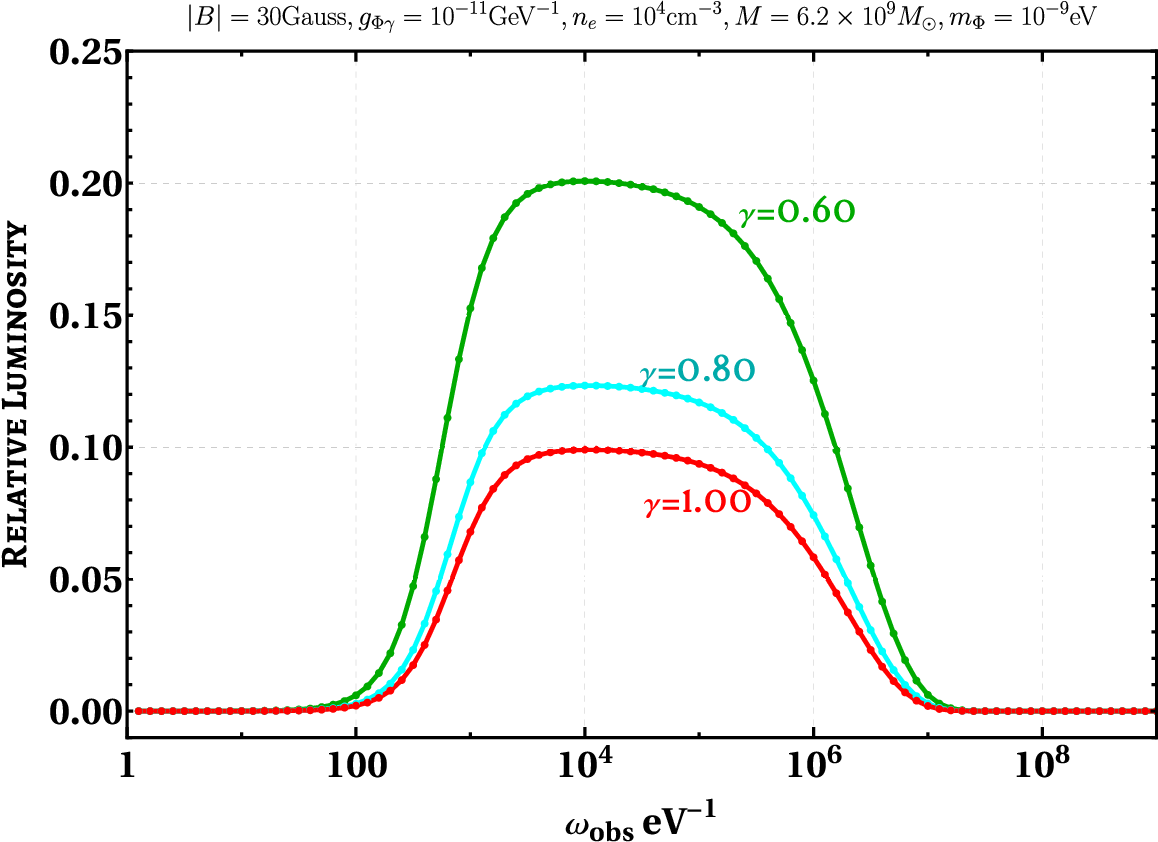}}\label{RLPH_c}}
\qquad
   \subfloat[\emph{The relative brightness spectrum of axions is shown with an axion mass of $m_{\Phi}=10~~\rm neV$, with varying values of the characteristic parameter $\gamma$.}]{{\includegraphics[width=7cm]{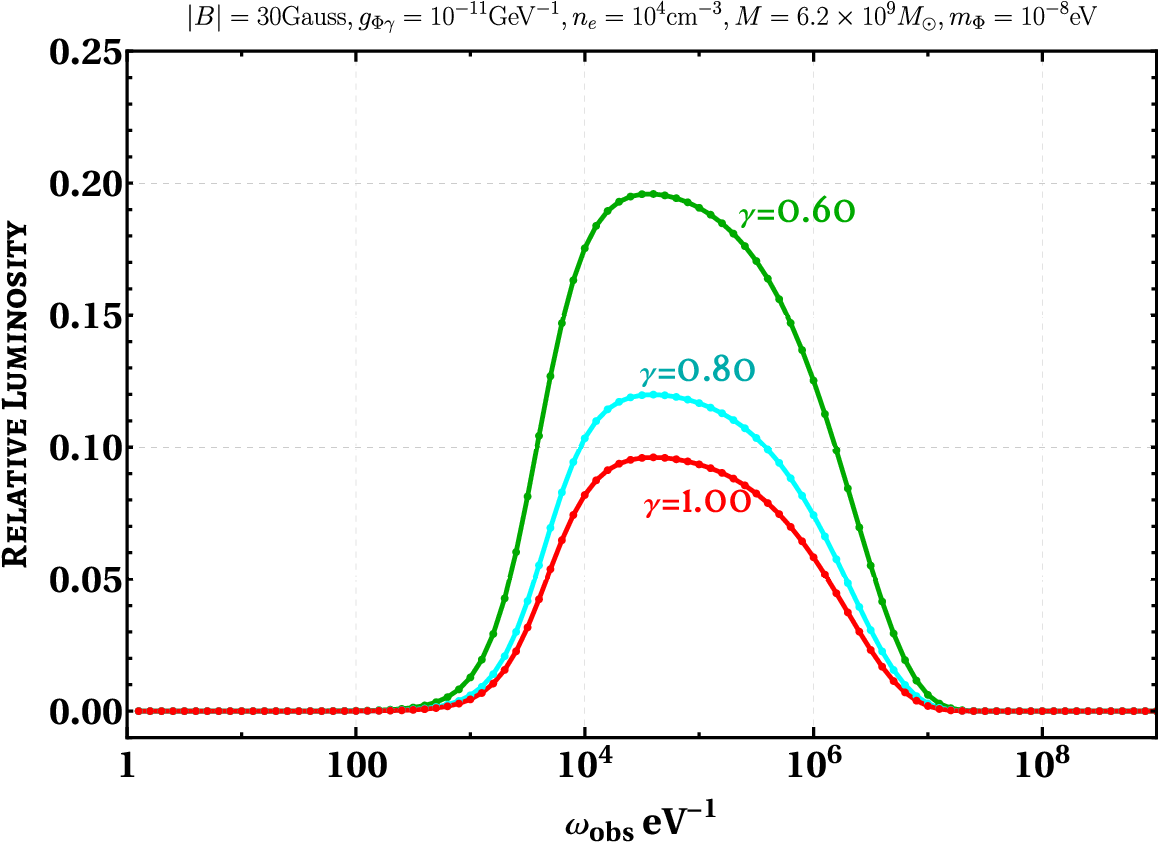}}\label{RLPH_d}}
    \caption{\emph{The relative luminosity spectra of photons has been depicted both before (using dotted lines) and after (using solid lines) the conversion of photons into axions. In this scenario, we have made the assumption that the magnetic field linked to the compact object is $|\mathbf{B}|=30 $ Gauss, the coupling between the axion and photon is $g_{\Phi\upgamma}=10^{-11}\rm  GeV^{-1}$, electron number density to be $n_{e}=10^{4} \rm cm^{-3}$ and the mass of the compact object is $M=6.2\times 10^{9} M_{\odot}$. Plots have been exhibited for two distinct mass values of axions. }}
    \label{Fig:Spectral_Var}
\end{figure}

In \ref{Fig:Spectral_Var}, we have illustrated the spectral variation of the relative luminosity for both photons and axions with respect to the characteristic parameter $\gamma$. The spectrum has been standardized based on the undistorted relative brightness for a Schwarzschild black hole at very low frequencies.  In \ref{RLPH_a} and \ref{RLPH_b}, the dotted lines show the relative luminosity of photons in the absence of photon conversion. While the solid lines show that the spectrum is reduced as a result of the fraction of photons becoming axions. The influence of the attenuation of the spectrum from its undistorted component becomes increasingly evident when the parameter is gradually reduced from its peak value $\gamma_{\rm max}=1$. From \ref{RLPH_a} and \ref{RLPH_b}, it can also be inferred that the percentage of attenuation along with initiation point  of attenuation-frequency depends on the mass of axions. In \ref{RLPH_c} and \ref{RLPH_d}, the variation of axion spectra has been depicted for a set of three values of characteristic parameter $\gamma$. It is noticed that the spread of the band of axion spectra for axion mass $m_{\Phi}=10 \rm ~neV$ is narrower than that of the spectrum for axion mass $m_{\Phi}=1\rm~ neV$.

\begin{figure}
    \centering
   \subfloat[\emph{Variation of relative luminosity of photons  has been depicted before the conversion of photons into axions.}]{{\includegraphics[width=7cm]{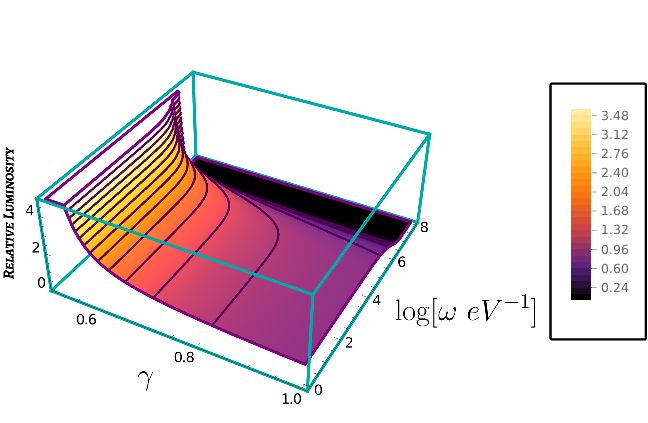}}\label{Fig:RLPH_a}}
   \qquad
   \subfloat[\emph{Variation of relative luminosity of photons  has been depicted after the conversion of photons into axions.}]{{\includegraphics[width=7cm]{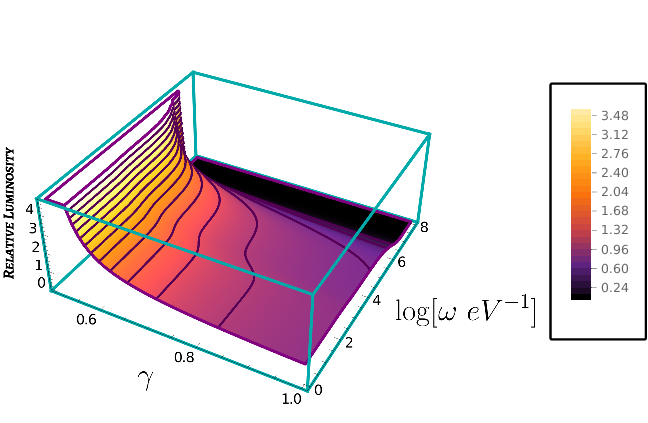}}\label{Fig:RLPH_b}}
   \qquad
   \subfloat[\emph{Variation of relative luminosity of axions produced via photon-to-axion conversion process.}]{{\includegraphics[width=7cm]{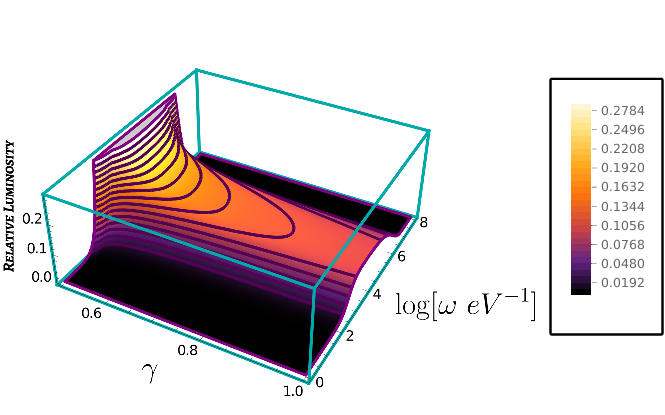}}\label{Fig:RLPH_c}}

    \caption{\emph{The relative luminosity spectra of photons and axions have been illustrated, with the state before and  after the conversion of photons into axions. It is assumed that the magnetic field connected to the compact object is $|\mathbf{B}|=30~~\rm Gauss$, the interaction between the axion and photon is $10^{-11}~~\rm  GeV^{-1}$, and the electron density is $n_{e}=10^{4}~~\rm cm^{3}$. The mass of the compact object is $M=6.2 \times  10^{9}  M_{\odot}$. The plots are illustrated for  axion mass value $m_{\Phi}=1~\rm  neV$.} }
    \label{Fig:CF_3D}
\end{figure}

In \ref{Fig:CF_3D} we have demonstrated the continuous variation of  relative luminosity spectrum with the simultaneous variation of characteristic parameter $\gamma$ in the specified range of $0.5<\gamma<1$, assuming axion mass to be $1~\rm neV$. In \ref{Fig:RLPH_a} and \ref{Fig:RLPH_b} we have depicted the variation of photon spectra without the photon-axion conversion and with photon-axion conversion, respectively. Axion spectrum arising due to the photon-axion conversion phenomena has been illustrated in \ref{Fig:RLPH_c}. This photon luminosity is suppressed at higher frequency because there are so few high-temperature electrons in the electromagnetic plasma that can produce photons at such high frequencies.

The condition of flat part of the photon spectrum can be understood with the help of \ref{eq:4.11}. The exponential factor, arising in the equation \ref{eq:4.11}, can be put to unity if
\begin{flalign}
    \dfrac{\omega_{e}}{T_{e}}=\dfrac{\omega_{c}}{T_{e,c}}\sqrt{\dfrac{f(r_{c})}{f(r_{e})}}\left(\dfrac{r_{e}}{r_{ph}} \right)\ll 1
\end{flalign}
In such circumstances, the \textit{Relative Luminosity (RL)} reaches its maximum value and the spectrum approaches an almost flat region.

\subsection{Required resolution of the image}
We expect that the photon-axion conversion will affect the observed photon spectrum. However, since the conversion occurs only near the photon sphere, we should note that only the spectrum near the photon sphere can be distorted. Thus, to observe the spectral distortion, we need to resolve the near-horizon region itself. Although the Event Horizon Telescope has effectively captured images of the structure near the compact object using radio waves, it is not currently equipped to perform high-resolution studies in the X-ray and gamma-ray ranges. Under these circumstances, the overall brightness emanating from the area surrounding the compact object will be significant. Total luminosity of photons emanating from a region $r_{\rm isco}<r_{e}<\mathscr{R}$ can be provided by the following approximate formula
\begin{flalign}
    \mathcal{L}_{\rm total} &=4\pi \int_{r_{\rm isco}}^{\mathscr{R}}\dd r_{e}~r_{e}^{2}\dfrac{dW}{\dd r_{e}\dd \omega_{e}\dd V_{e} } \sim \dfrac{16}{\epsilon^{2}}\left(\dfrac{r_{ph}}{M} \right)^{3}\left[\sqrt{\left(\dfrac{\mathscr{R}}{r_{ph}} \right)}-\sqrt{\left(\dfrac{r_{\rm isco}}{r_{ph}} \right)}  \right]\mathcal{L}_{\omega}^{0}\label{Eq:TT}
\end{flalign}
where $r_{\rm isco}$ represents the radius of innermost stable circular orbit. From \ref{Eq:TT}, one can conclude that the majority of the overall brightness will be attributed to the emission originating from the area beyond photon sphere. As a result, the reduction in brightness caused by the conversion will be negligible. We know that the Chandra observatory has angular resolution of order of arc-sec and hence unable to detect the dimming phenomena for M87* since it observes the area beyond the photon sphere. The observation of such dimming can be possible only if we increase the resolution of the captured image of the compact object. To achieve such precision, we need to track the photons near the photon sphere of the compact object, i.e, over a region where $\mathscr{R}\sim r_{ph}$. The required resolution for such scenario is
\begin{flalign}
    \theta=\eval{\dfrac{\mathscr{R}}{D}}_{\gamma=1}\lesssim 10^{-5} {\rm arcsec} \left(\dfrac{M}{6.2\times 10^{9}M_{\odot}} \right)\left(\dfrac{16.8 {\rm Mpc}}{D} \right)
\end{flalign}
 
where $D$ is the distance of compact object from the observer. So, for M87* compact object,  even for X-ray and gamma-ray, we need the angular resolution $\theta\lesssim 10 ~\upmu{\rm as}$. In \ref{Fig:Spectral_Var} and \ref{Fig:CF_3D}, The expected energy spectrum are illustrated with multiple examples. The horizontal axis represent the frequency we observe, denoted as $\omega_{\rm obs}$. This frequency is connected to the frequency at the photon sphere, denoted as $\omega_{c}$, by the equation $\omega_{\rm obs}=\sqrt{f(r_{ph})}\omega_{c}$, where $\sqrt{f(r_{ph})}$ is a constant factor accounting for gravitational redshift. We disregarded further minor influences, such as peculiar velocities and cosmic expansion. In each of such scenarios, the spectral luminosity has been normalised by the infrared value for $\gamma=1$ case. We also assumed that the photons are created by thermal bremsstrahlung of the gas spread over a spherical region (starting from $r_{\rm isco}$)  around  the compact object. One should remember that the cut-off frequency of  the spectrum is at around $\omega_{\rm obs}\sim T_{e,c}\sim 10^{7} ~\rm eV$ due to the exponential suppression factor in \ref {Eq:Suppression}. For $\theta\simeq 1 ~\rm arcsec$, due to the lack of angular resolution, we are unable to discover any spectral distortion in the X-ray spectrum. On the other hand, when $\theta$ equals around few $\rm \upmu as$, we are able to resolve the photon sphere itself, which allows us to observe the reduction in the brightness for the X-ray spectrum.

{Detecting photon-axion conversion at high frequencies, such as in the X-ray regime, presents significant observational challenges. One of the primary limitations is the reduced flux density of photon rings at these frequencies, particularly for higher-order images, which are exceedingly faint and may fall below the sensitivity thresholds of current observational facilities. Additionally, the optical depth of X-rays in the strong magnetic fields near compact objects remains uncertain, further complicating the detectability of the effect. This limitation arises due to the complex interplay between the absorption, scattering, and conversion processes in the surrounding plasma environment. Addressing these challenges would require substantial advancements in observational technologies, such as higher sensitivity instruments and enhanced resolution capabilities in the X-ray regime. Furthermore, complementary strategies, such as correlating multi-wavelength observations and incorporating alternative indirect signatures of axion-like particles, could help overcome these barriers. Future missions equipped with next-generation X-ray observatories may offer a more promising avenue to probe such subtle effects and validate the theoretical predictions presented in this work.}

\section{Conclusion}

{ 
This study examines photon-axion conversion within the Janis-Newman-Winicour (JNW) naked singularity spacetime, concentrating on the attenuation of the photon ring at high frequencies. Through the examination of photon trajectories and their interaction with axions in a strong magnetic field, we established that this conversion process may result in diminished luminosity of the observed photon ring, especially within the X-ray and gamma-ray spectra. The frequency-dependent dimming effect, if detected, may offer an indirect indication of axion-like particles (ALPs) in astrophysical contexts. We calculated the photon-axion conversion rate assuming a constant magnetic field and utilized this to assess the consequent dimming effect on the luminance of the photon ring. The conversion process is contingent upon frequency, with dimming becoming more significant in the X-ray and gamma-ray regions. Our calculations suggest that this dimming, if detected, may act as a unique signature of axion-like particles (ALPs) near naked singularities. The identification of such a signature would not only offer indirect proof for the existence of ALPs but would also aid in differentiating naked singularities from black holes, which possess distinct photon ring patterns owing to the presence of an event horizon.

The   key novelty of this work  is the extension of photon-axion conversion research to a horizonless compact object, namely the JNW naked singularity. Although previous research has predominantly focused on black hole spacetimes, we have shown that photon-axion conversion can transpire in the more unconventional setting of a naked singularity, resulting in unique observable signs. The photon ring surrounding a naked singularity is significant due to the absence of an event horizon, indicating that photon paths and conversion processes may display distinctive characteristics that can be utilized to investigate the properties of the underlying spacetime. This research enhances the understanding of axion-photon interactions and establishes a novel framework for investigating horizonless spacetimes in astrophysical contexts. Moreover, our findings present an innovative approach for investigating the characteristics of ALPs and  structure of compact objects. The capability to identify photon ring dimming resulting from photon-axion conversion has intriguing prospects for forthcoming measurements. The existing observational capabilities, exemplified by the Event Horizon Telescope (EHT), predominantly concentrate on lower-frequency emissions and are now unable to detect high-frequency events where photon-axion conversion is most pronounced. The forthcoming generation of observational devices, such as the next-generation Event Horizon Telescope (ngEHT)\cite{Deliyski:2024wmt},  GRAVITY collaboration, future space-based observatory guarantees enhanced resolution and augmented sensitivity to X-ray and gamma-ray emissions. These advancements will enable the observation of the anticipated high-frequency dimming, facilitating the direct evaluation of the theoretical framework we have established.

The observational signatures examined in this study, including the frequency-dependent attenuation of the photon ring, signify a significant progression in the quest for axion-like particles and alternative compact objects.
But there are well-known physical processes, such as synchrotron self-absorption, bremsstrahlung, and inverse Compton scattering, which cause high-frequency emissions to naturally weaken as frequency increases. This raises the question of how to distinguish natural dimming from the dimming caused by photon-axion conversion. However there are ways to differentiate the reason of the dimming.  Natural dimming mechanisms typically follow well-established patterns, which are determined by the emission mechanism and the properties of the medium (e.g., plasma density, temperature, magnetic field strength). In contrast, photon-axion conversion would introduce a distinct frequency-dependent effect, specifically tied to the axion-photon coupling and the strength of the magnetic field. The conversion efficiency varies with frequency and distance from the central object in a manner that could, in principle, be distinguished from other dimming effects. Photon-axion conversion is strongly dependent on the magnetic field strength, axion mass, and the coupling constant. In regions where photon-axion conversion is efficient, the dimming should exhibit a sharp frequency dependence that deviates from the more gradual dimming caused by thermal or synchrotron effects. In short,  the axion-induced dimming would leave a specific spectral signature that, with careful analysis, could be distinguished from natural dimming. Future high-precision observations could allow for this distinction by comparing the predicted and observed frequency-dependent luminosity profiles. Photon-axion conversion affects only photons with specific polarization, meaning that an additional observational signature could be detected by looking at the polarization of the light. Astrophysical sources typically exhibit polarized emissions, and photon-axion conversion could induce a characteristic change in the polarization spectrum that would further differentiate it from natural dimming mechanisms.

This paper emphasizes the significance of examining alternative compact objects beyond black holes, alongside its observational potential. Naked singularities, though conjectural, question our comprehension of gravitational collapse and the cosmic censorship conjecture. }There are various naked singularity in literature. Out of these, the most important naked singularity solution is Janis-Newman-Winicour spacetime, which has been considered here for our analysis. For the sake of simplicity, in our analysis we have assumed that the magnetic field and plasma density are uniform in the vicinity of the photon sphere. The sources that release photons through a radiative bremsstrahlung process are believed to exist outside the photon sphere. The brightness is mostly influenced by the lower limit of the integral in  \ref{eq:4.9} rather than the upper limit. Therefore, the distance we choose from the centre of the compact object is of greater significance. In this study, we have defined the lower limit as the innermost stable circular orbit (ISCO). Another important thing to notice is that the “scattering-limit-factor” ($\mathfrak{S}$)  as defined in \ref{Eq:RQ} has an upper limit $\mathfrak{S}_{\rm upp}=3$ and monotonically decreases to $\mathfrak{S}_{\rm low}=0$ as we vary the characteristic parameter $\gamma$ of JNW spacetime from $\gamma=1$ to $\gamma=0.5$. So the impact of the effect of scattering of photons to determine the conversion factor is more prominent for the Schwarzschild black hole than that of the JNW naked singularity. In fact, for JNW spacetime with characteristic parameter $\gamma$ very near to limiting value $\gamma_{\rm low}=0.5$, one can safely ignore the effect of scattering phenomena entirely, even for very small value of photon mean free path.

 By exploring photon-axion conversion in the JNW spacetime, we offer a novel perspective on distinguishing horizonless objects from black holes through their photon ring structures and corresponding electromagnetic signatures. This contributes a crucial aspect to the ongoing endeavors to evaluate general relativity in strong-field contexts and to investigate the potential existence of exotic compact objects that lie outside conventional black hole physics. In summary, this research of photon-axion conversion in the JNW naked singularity spacetime provides new insights into its observable effects.The dimming of the photon ring at high frequencies, caused by photon-axion interactions, provides a unique signature that could be detected with future high-resolution telescopes.  Our findings improve our understanding of photon-axion interactions in extreme astrophysical conditions and help distinguish black holes from horizonless objects. The framework we established in this publication could help find axion-like particles and explore alternate gravitational structures as observational techniques improve.

There are other avenues to explore beyond the current project. One approach is to incorporate the rotational motion of the compact object. The rotating compact object exhibits the intriguing phenomenon of photons orbiting at two distinct radii on the equatorial plane. Due to the gravitational redshift experienced by photons released from various distances, the conversion into axions at those radii may result in dimming at different frequencies. It is also crucial to investigate the impact of photon-axion conversion on the polarisation of light emitted from the photon sphere. It is also worth studying conversion not only in the background of the magnetic field but also in the background of the axion\cite{Masaki:2019ggg,Dror:2021nyr}, since they are the potential candidate for dark matter and may be produced due to superradiance instability of the compact object \cite{Brito:2015oca,Day:2019bbh}. Another possible direction is to test the effect of conversion mechanism for the regular compact objects and even for regularised version of JNW spacetime\cite{Pal:2022cxb} and to explore astrophysical aspects and the possibility of distinguishing such solutions from singular ones.

\section*{Acknowledgement}
SS is grateful to Prof. Joseph Patrick Conlon for providing useful tips and reviewing the manuscript. We are grateful to the anonymous referees for their fruitful comments and suggestions that lead to a concise and better  version of the manuscript.

{

\appendix
\renewcommand{\theequation}{A\arabic{equation}}
\setcounter{equation}{0}
\section*{Appendix A: Photons approaching the photon sphere}\label{SEC:PAPS}
\subsection{Connection between impact parameter and emission angle}
Let us consider a beam of light directed towards the photon sphere of a compact object with an impact parameter $b$ from a point $p_e$ located at $r_e$ in spherically symmetric spacetime coordinates. Introducing an angle $\Upsilon_e$ between the initial direction of the incident photon and the line toward the centre of the compact object, as shown in \ref{fig2}, allows us to characterize the trajectory of the photon. 
\begin{figure}[htbp]
	\centering
	\includegraphics[scale=0.3]{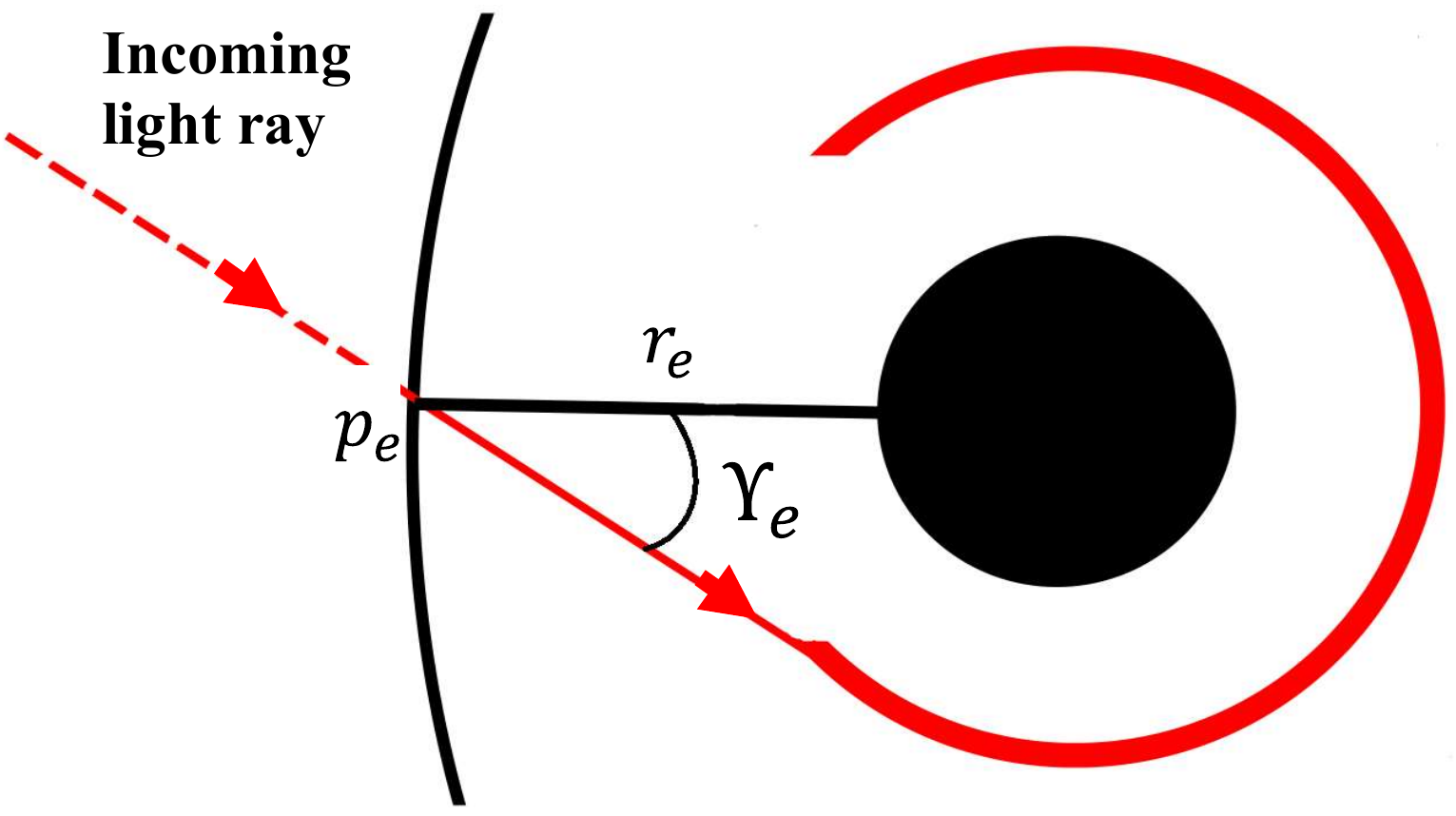}
	\caption{\emph{The trajectory of the light ray originating from a point $p_e$ towards the photon sphere is depicted by the red curve. This point, $p_e$, is situated on a sphere with a radius denoted as $r_e$, centred around the compact object within a spherically symmetric coordinate system. The angle $\Upsilon_e$ corresponds to the zenith angle.}}
	\label{fig2}
\end{figure}

In the context of the spherically symmetric spacetime described by the \ref{Eq:SSM1} and assuming $g(r)=f(r)$, the tetrad vectors are provided by
\begin{subequations}
	\begin{align}
		& e^{(0)}=\left(\sqrt{f(r)}, 0,0,0\right), \\
		& e^{(1)}=\left(0, \frac{1}{\sqrt{f(r)}}, 0,0\right), \\
		& e^{(2)}=\left(0,0, r \mathcal{R}(r), 0\right), \\
		& e^{(3)}=\left(0,0,0, r \mathcal{R}(r)\sin \theta\right).
	\end{align}
\end{subequations}
The tetrads $e^{(1)}$ and $e^{(3)}$  serve as orthonormal bases, aligning parallel and perpendicular to the path leading to the centre of compact objects, respectively. Consequently, the angle $\Upsilon_e$ is expressed as:
\begin{equation}
	\tan \Upsilon_e=\left|\frac{k^\mu e_\mu^{(3)}}{k^\mu e_\mu^{(1)}}\right|_{p_e}=\left|r\mathcal{R}(r) \sqrt{f(r)} \sin \theta \frac{d \phi}{d r}\right|_{p_e},
\end{equation}
here, $k^\mu={d x^\mu}/{d \lambda}$ represents the tangent vector tracing the geodesics of the photons with the affine parameter $\lambda$. For simplicity, we assume that the geodesic plane lies in the $\theta=\pi / 2$ plane. In such scenario, one can derive
\begin{equation}
	b=\frac{r_e\mathcal{R}(r_{e})}{\sqrt{f\left(r_e\right)}} \sin \Upsilon_e.\label{eq:4.3}
\end{equation}
This equation establishes a link between the emission angle $\Upsilon_e$,  evaluated at $p_e$, and the impact parameter $b$.

\subsection{Photon sphere inflow from spherical region}
We envision the compact object situated at the centre of a sphere, where photons are emitted isotropically from every point with a specific emissivity. We aim to estimate the approximate number of photons approaching a photon sphere. To accomplish this, we adopt a spherically symmetric spacetime as our model geometry. {\bf We chose the spherical, isotropic accretion flow model as a simplifying assumption to make the analysis tractable\cite{10.1093/mnras/sty2624,Narayan:2019imo,Bambi:2013nla}. Realistic accretion flows may indeed be more complex and anisotropic, but our aim is to provide an initial exploration of how photon-axion conversion might operate in this simplified scenario.}

Let $\dd N/\dd \tau_e$ represent the number of photons emitted with a frequency width $d\omega_e$ from $dV_e$, and travelling through an infinitesimal solid angle $d\omega_e$ per unit time $\tau_e$ as observed from the emission point $p_e$. Here, $V_e$, $\omega_e$, $\tau_e$, and $\Omega$ are all in a local inertial frame at $p_e$, as shown in  \ref{fig3}.
\begin{figure}[ht]
	\centering
	\includegraphics[scale=0.38]{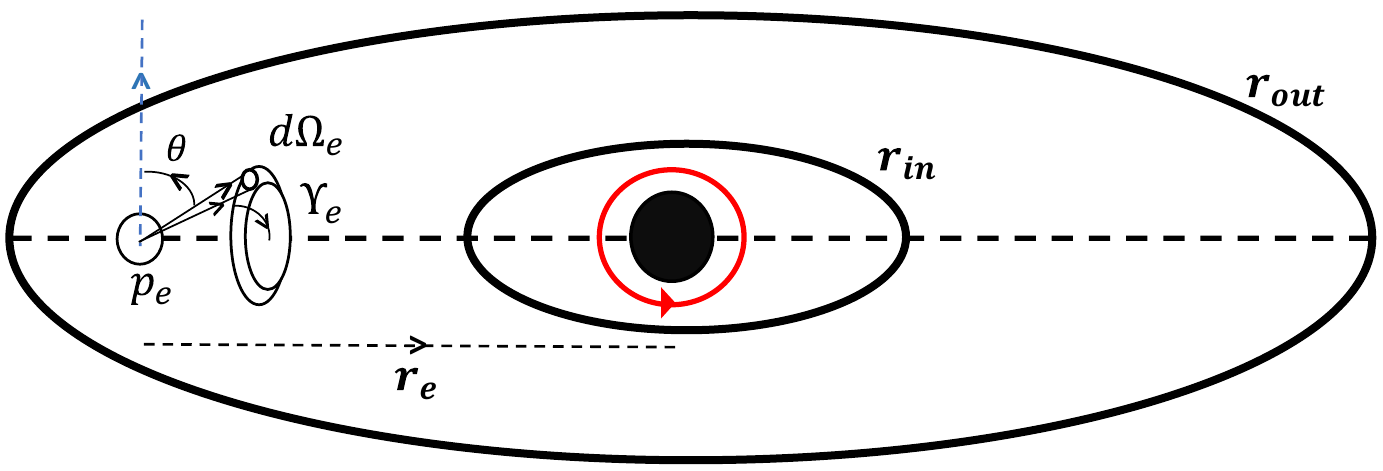}
	\caption{\emph{Photons are emitted from a spherical region with boundaries defined by $r_{in}$ and $r_{out}$, and they converge towards the photon sphere of the compact object, which is represented by the red circle.}}\label{fig3}
\end{figure}

Designating $\Psi_e$ as the azimuthal angle in the plane perpendicular to the direction from point $p_e$ to the compact object, and $\Upsilon_e$ as the zenithal angle measured from that direction, we assume isotropic emission from $p_e$. Under this assumption, we consider $J_e$ to be independent of $\Upsilon_e$ and $\Psi_e$. Hence, the expression for a photon with all the conditions is given as
\begin{equation}
	\dd^6\left(\frac{\dd N}{\dd \tau_e}\right)=J_e\left(\omega_e, r_e\right) \dd \Omega_e \dd V_e \dd \omega_e.\label{eq:4.4}
\end{equation}
We reframe  \ref{eq:4.4} in relation to the impact parameter $b$ of a photon. Employing  \ref{eq:4.3} and $\dd b=\dfrac{r_e \mathcal{R}(r_{e}}{\sqrt{f(r_e)}}\cos{\Upsilon_e}d\Upsilon_e$, we obtain the intended outcome for a constant $r_e$. Upon integration \ref{eq:4.4} across the angle $\Psi_e$, we determine the number of photons emitted towards the direction of the photon sphere  per unit time as
\begin{flalign}
	\dd^5\left(\frac{\dd N}{\dd \tau_e}\right)=\frac{1}{2} \times 2 \pi J_e\left(\omega_e, r_e\right) \dfrac{\sqrt{f\left(r_e\right)}}{r_e\mathcal{R}(r_{e})} \dfrac{b}{\sqrt{\dfrac{r_e^2\mathcal{R}^{2}(r_{e})}{f\left(r_e\right)}-b^2}}\times \dd b \,\dd V_e \dd \omega_e \label{eq:4.5}
\end{flalign}
To account for the restriction that only photons within  angle of $0 \leq \Upsilon \leq \pi/2$ can approach the photon sphere, we multiply by a factor of $1/2$ in \ref{eq:4.5}. Within a local inertial frame at $p_e$, with volume element $\dd V_e=(r_e^2\mathcal{R}^{2}(r_{e})/\sqrt{f(r_e)})\sin{\theta}\dd r_e\,\dd\theta \dd\phi$ and the time element $\dd\tau_e=\sqrt{f(r_e)}\dd t$ in spherically symmetric spacetime, the number count of photons with an impact parameter falling within the range $(b,b+\dd b)$, emanating from a spherical shell with a width of $\dd r_e$ and a unit frequency of $\omega_e$, is determined by integrating \ref{eq:4.5} over $\theta$ and $\phi$, given as
\begin{equation}
	\frac{\dd^4 N}{\dd t \dd \omega_e \dd b \dd r_e}=4 \pi^2 J_e\left(\omega_e, r_e\right) \frac{b r_e \mathcal{R}(r_{e})\sqrt{f\left(r_e\right)}}{\sqrt{\dfrac{r_e^2 \mathcal{R}^{2}(r_{e})}{ f\left(r_e\right)}-b^2}}.\label{eq:4.6}
\end{equation}
It is important to note that $\omega_e$ represents the frequency in a local inertial frame at the emission point $p_e$, and it is defined as
\begin{equation}
	\omega_e=\left.k^\mu e_\mu^{(0)}\right|_{p_e}=\left.\frac{d t}{d \lambda} \sqrt{f\left(r_e\right)}\right|_{p_e}=\frac{E}{\sqrt{f\left(r_e\right)}},\label{eq:4.7}
\end{equation}
here, $k^\mu=\dd x^\mu/\dd\lambda$ represents the tangent vector to the geodesic, $\lambda$ denotes the affine parameter, and $e^{(\alpha)}_\mu$ signifies the local tetrad at the emission point $p_e$. The frequency measured in a local inertial frame at the photon sphere situated at $r=r_{{ph}}$ is denoted as $\omega_c\equiv \dfrac{E}{\sqrt{f(r_{ph})}}$.  With an impact parameter in the range $(b,b+db)$, the quantity of photons that reach the photon sphere within a specific time interval $dt$ and frequency $\omega_{c}$ is governed by
\begin{equation}
	\left(\frac{\dd^3 N}{\dd t \dd \omega_c \dd b}\right)=4 \pi^2 \int_{r_{\text {in }}}^{r_{\text {out }}} \mathrm{d} r_e J_e\left(\frac{\sqrt{f\left(r_{{ph}}\right)} \omega_c}{\sqrt{f\left(r_e\right)}}, r_e\right)\times \dfrac{b r_e \mathcal{R}(r_{e})\sqrt{f\left(r_e\right)}}{\sqrt{\dfrac{r_e^2\mathcal{R}^{2}(r_{e}) }{ f\left(r_e\right)}-b^2}}.\label{eq:4.8}
\end{equation}
In this context, the emission region is confined within a spherical region with an inner radius $r_{\mathrm{in}}$
and an outer radius $r_{\mathrm{out}}$. Given our focus on the vicinity of the photon sphere, integrating \ref{eq:4.8} within the range ($b_c,b_c(1+\mathfrak{a}\epsilon^2)$), we obtain
\begin{equation}
	\frac{\dd^2 N}{\dd t \dd \omega_c} \simeq 4 \pi^2 \mathfrak{a} \epsilon^2 b_c^2 \int_{r_{\text {in }}}^{r_{\text {out }}} \mathrm{d} r_e J_e\left(\frac{\sqrt{f\left(r_{\mathrm{ph}}\right)} \omega_c}{\sqrt{f\left(r_e\right)}}, r_e\right) \times \frac{r_e \mathcal{R}(r_{e})\sqrt{f\left(r_e\right)}}{\sqrt{\dfrac{r_e^2\mathcal{R}^{2}(r_{e}) }{ f\left(r_e\right)}-b_c^2}},\label{eq:4.9}
\end{equation}
Here, it is presumed that $r_{e}$ is significantly greater than the radius of the photon sphere, such that $r_{\text{in}}^2\mathcal{R}^{2}(r_{\rm in})/f(r_{\text{in}})>b^2$ holds for $b$ in the range ($b_c,b_c(1+\mathfrak{a}\epsilon^2)$).

\subsection{Photonic sources in the proximity to galactic compact objects}
Recent observations have revealed the presence of an electromagnetic plasma in the vicinity of a supermassive compact object \cite{EventHorizonTelescope:2019pgp}. Photons with frequencies significantly higher than the plasma frequency ($\omega \gg \omega_{pl}$) can traverse this plasma. This high-frequency radiation originates from charged particles being accelerated within the Coulomb field of another charge, a phenomenon commonly referred to as free-free emission or bremsstrahlung. Understanding this process thoroughly requires a quantum treatment, as it enables the production of photons with energies comparable to those of the emitting particles. The energy emitted by this process per unit time, per unit frequency, and unit volume is mathematically represented as \cite{rybicki2004radiative}
\begin{equation}
	\frac{\dd W}{
		\dd \tau_e 
		\dd \omega_e \dd V_e}=\frac{2^4 \alpha^3}{3 m_e}\left(\frac{2 \pi}{3 m_e}\right)^{1 / 2} T_e^{-1 / 2} n_e^2 e^{-\omega_e / T_e} \bar{g}_{f f},\label{eq:4.10}
\end{equation}
here, $T_e$ represents the electron temperature, $n_e$ denotes the electron number density in the plasma, and $\bar{g}_{ff}$ represents the velocity-averaged Gaunt factor for free emission. The emission rate is expressed as an approximate classical result multiplied by the free emission Gaunt factor $\bar{g}_{ff}$, which takes into account the quantum-mechanical Born approximation. Although $\bar{g}_{ff}$ is a function of the energy of the electron and the frequency of emission, for order-of-magnitude estimation, can be considered to be approximately unity. Additionally, we assume that the ion density $n_i$ is equal to $n_e$. For isotropic radiation, $J_e^{(N)}$ as defined in \ref{eq:4.4} gives
\begin{equation}
	J_e^{(N)}\left(\omega_e, r_e\right)=\frac{1}{4 \pi \omega_e}\left(\frac{2^4 \alpha^3}{3 m_e}\right)\left(\frac{2 \pi}{3 m_e}\right)^{1 / 2} \times T_e^{-1 / 2} n_e^2 e^{-\omega_e / T_e} \bar{g}_{f f}.\label{eq:4.11}
\end{equation}
The Event Horizon Telescope (EHT) images of M87* reveal a bright ring-like structure, corresponding to the emission from the inner part of the hot accretion disk. The observed properties of the emission, such as its spectrum and variability, align with expectations from a hot accretion flow. The temperature of hot accretion is nearly virial, approximately $T \approx G M m_p/6k_Br \approx  10^{12}K(r/M)^{-1}$, with $m_p$ representing the proton mass \cite{Quataert:2002xn,Yuan:2014gma}. When studying accretion processes around supermassive compact objects like M87*, theoretical models are commonly employed to understand the properties and behaviour of the accreting material. One such model is the spherical accretion model, which assumes that the accretion flow onto the compact object is spherically symmetric. However, it's important to note that while the spherical accretion model serves as a useful starting point for understanding the overall behaviour of the accretion flow around M87*, it simplifies the actual complex processes that may occur. The spherical mass accretion rate can be expressed as $M = 4\pi r^2\rho v_r$, with mass density $\rho$ and radial velocity $v_r$. Assuming constant mass accretion and free-falling gas, $vr \propto r^{-1/2}$, we find that $\rho \propto (r/M)^{-3/2}$. Consequently, we can assume that the electron temperature and number density of electrons obey a power law
\begin{subequations}
	\begin{align}
		& T_e=T_{e, c}\left(\frac{r_e}{r_{\mathrm{ph}}}\right)^{-1},\\
		& n_e=n_{e, c}\left(\frac{r_e}{r_{\mathrm{ph}}}\right)^{-3 / 2},
	\end{align}
\end{subequations}
where $T_{e,c}$ and $n_{e,c}$ denote the value at the photon sphere.

}

\bibliography{references}

\providecommand{\href}[2]{#2}\begingroup\raggedright\begin{thebibliography}{100}

\bibitem{peccei1977cp}
R.~D. Peccei and H.~R. Quinn, ``Cp conservation in the presence of
  instantons,'' {\em Phys. Rev. Lett} {\bfseries 38} no.~328, (1977)
  1440--1443.

\bibitem{weinberg1978new}
S.~Weinberg, ``A new light boson?,'' {\em Physical Review Letters} {\bfseries
  40} no.~4, (1978) 223.

\bibitem{wilczek1978problem}
F.~Wilczek, ``Problem of strong p and t invariance in the presence of
  instantons,'' {\em Physical Review Letters} {\bfseries 40} no.~5, (1978) 279.

\bibitem{kim1979weak}
J.~E. Kim, ``Weak-interaction singlet and strong cp invariance,'' {\em Physical
  Review Letters} {\bfseries 43} no.~2, (1979) 103.

\bibitem{shifman1980can}
M.~A. Shifman, A.~Vainshtein, and V.~I. Zakharov, ``Can confinement ensure
  natural cp invariance of strong interactions?,'' {\em Nuclear Physics B}
  {\bfseries 166} no.~3, (1980) 493--506.

\bibitem{dine1981simple}
M.~Dine, W.~Fischler, and M.~Srednicki, ``A simple solution to the strong cp
  problem with a harmless axion,'' {\em Physics letters B} {\bfseries 104}
  no.~3, (1981) 199--202.

\bibitem{zhitnitskii1980possible}
A.~R. Zhitnitsky, ``{On Possible Suppression of the Axion Hadron Interactions.
  (In Russian)},'' {\em Sov. J. Nucl. Phys.} {\bfseries 31} (1980) 260.

\bibitem{Conlon:2006tq}
J.~P. Conlon, ``{The QCD axion and moduli stabilisation},''
  \href{http://dx.doi.org/10.1088/1126-6708/2006/05/078}{{\em JHEP} {\bfseries
  05} (2006) 078}, \href{http://arxiv.org/abs/hep-th/0602233}{{\ttfamily
  arXiv:hep-th/0602233}}.

\bibitem{armengaud2014conceptual}
E.~Armengaud, F.~Avignone, M.~Betz, P.~Brax, P.~Brun, G.~Cantatore, J.~Carmona,
  G.~Carosi, F.~Caspers, S.~Caspi, {\em et~al.}, ``Conceptual design of the
  international axion observatory (iaxo),'' {\em Journal of Instrumentation}
  {\bfseries 9} no.~05, (2014) T05002.

\bibitem{cast2017new}
C.~collaboration, ``New cast limit on the axion--photon interaction,'' {\em
  Nature Physics} {\bfseries 13} no.~6, (2017) 584--590.

\bibitem{asztalos2010squid}
S.~J. Asztalos, G.~Carosi, C.~Hagmann, D.~Kinion, K.~Van~Bibber, M.~Hotz,
  L.~Rosenberg, G.~Rybka, J.~Hoskins, J.~Hwang, {\em et~al.}, ``Squid-based
  microwave cavity search for dark-matter axions,'' {\em Physical review
  letters} {\bfseries 104} no.~4, (2010) 041301.

\bibitem{csaki2002dimming}
C.~Csaki, N.~Kaloper, and J.~Terning, ``Dimming supernovae without cosmic
  acceleration,'' {\em Physical Review Letters} {\bfseries 88} no.~16, (2002)
  161302.

\bibitem{csaki2002effects}
C.~Csaki, N.~Kaloper, and J.~Terning, ``Effects of the intergalactic plasma on
  supernova dimming via photon--axion oscillations,'' {\em Physics Letters B}
  {\bfseries 535} no.~1-4, (2002) 33--36.

\bibitem{deffayet2002dimming}
C.~Deffayet, D.~Harari, J.-P. Uzan, and M.~Zaldarriaga, ``Dimming of supernovae
  by photon-pseudoscalar conversion and the intergalactic plasma,'' {\em
  Physical Review D} {\bfseries 66} no.~4, (2002) 043517.

\bibitem{grossman2002effects}
Y.~Grossman, S.~Roy, and J.~Zupan, ``Effects of initial axion production and
  photon--axion oscillation on type ia supernova dimming,'' {\em Physics
  Letters B} {\bfseries 543} no.~1-2, (2002) 23--28.

\bibitem{galanti2023observability}
G.~Galanti, L.~Nava, M.~Roncadelli, F.~Tavecchio, and G.~Bonnoli,
  ``Observability of the very-high-energy emission from grb 221009a,'' {\em
  Physical Review Letters} {\bfseries 131} no.~25, (2023) 251001.

\bibitem{Conlon:2014xsa}
J.~P. Conlon and F.~V. Day, ``{3.55 keV photon lines from axion to photon
  conversion in the Milky Way and M31},''
  \href{http://dx.doi.org/10.1088/1475-7516/2014/11/033}{{\em JCAP} {\bfseries
  11} (2014) 033}, \href{http://arxiv.org/abs/1404.7741}{{\ttfamily
  arXiv:1404.7741 [hep-ph]}}.

\bibitem{troitsky2022parameters}
S.~V. Troitsky, ``Parameters of axion-like particles required to explain
  high-energy photons from grb 221009a,'' {\em JETP Letters} {\bfseries 116}
  no.~11, (2022) 767--770.

\bibitem{baktash2022interpretation}
A.~Baktash, D.~Horns, and M.~Meyer, ``{Interpretation of multi-TeV photons from
  GRB221009A},'' \href{http://arxiv.org/abs/2210.07172}{{\ttfamily
  arXiv:2210.07172 [astro-ph.HE]}}.

\bibitem{lin2023electroweak}
W.~Lin and T.~T. Yanagida, ``Electroweak axion in light of grb221009a,'' {\em
  Chinese Physics Letters} {\bfseries 40} no.~6, (2023) 069801.

\bibitem{gonzalez2023grb}
M.~Gonzalez, D.~A. Rojas, A.~Pratts, S.~Hernandez-Cadena, N.~Fraija, R.~Alfaro,
  Y.~P. Araujo, and J.~Montes, ``Grb 221009a: A light dark matter burst or an
  extremely bright inverse compton component?,'' {\em The Astrophysical
  Journal} {\bfseries 944} no.~2, (2023) 178.

\bibitem{nakagawa2023axion}
S.~Nakagawa, F.~Takahashi, M.~Yamada, and W.~Yin, ``Axion dark matter from
  first-order phase transition, and very high energy photons from grb
  221009a,'' {\em Physics Letters B} {\bfseries 839} (2023) 137824.

\bibitem{carenza2022alp}
P.~Carenza and M.~C.~D. Marsh, ``{On ALP scenarios and GRB 221009A},''
  \href{http://arxiv.org/abs/2211.02010}{{\ttfamily arXiv:2211.02010
  [astro-ph.HE]}}.

\bibitem{wang2023axion}
L.~Wang and B.-Q. Ma, ``Axion-photon conversion of grb221009a,'' {\em Physical
  Review D} {\bfseries 108} no.~2, (2023) 023002.

\bibitem{Conlon:2015uwa}
J.~P. Conlon, M.~C.~D. Marsh, and A.~J. Powell, ``{Galaxy cluster thermal x-ray
  spectra constrain axionlike particles},''
  \href{http://dx.doi.org/10.1103/PhysRevD.93.123526}{{\em Phys. Rev. D}
  {\bfseries 93} no.~12, (2016) 123526},
  \href{http://arxiv.org/abs/1509.06748}{{\ttfamily arXiv:1509.06748
  [hep-ph]}}.

\bibitem{Berg:2016ese}
M.~Berg, J.~P. Conlon, F.~Day, N.~Jennings, S.~Krippendorf, A.~J. Powell, and
  M.~Rummel, ``{Constraints on Axion-Like Particles from X-ray Observations of
  NGC1275},'' \href{http://dx.doi.org/10.3847/1538-4357/aa8b16}{{\em Astrophys.
  J.} {\bfseries 847} no.~2, (2017) 101},
  \href{http://arxiv.org/abs/1605.01043}{{\ttfamily arXiv:1605.01043
  [astro-ph.HE]}}.

\bibitem{Conlon:2017qcw}
J.~P. Conlon, F.~Day, N.~Jennings, S.~Krippendorf, and M.~Rummel,
  ``{Constraints on Axion-Like Particles from Non-Observation of Spectral
  Modulations for X-ray Point Sources},''
  \href{http://dx.doi.org/10.1088/1475-7516/2017/07/005}{{\em JCAP} {\bfseries
  07} (2017) 005}, \href{http://arxiv.org/abs/1704.05256}{{\ttfamily
  arXiv:1704.05256 [astro-ph.HE]}}.

\bibitem{hooper2007detecting}
D.~Hooper and P.~D. Serpico, ``Detecting axionlike particles with gamma ray
  telescopes,'' {\em Physical Review Letters} {\bfseries 99} no.~23, (2007)
  231102.

\bibitem{hochmuth2007effects}
K.~A. Hochmuth and G.~Sigl, ``Effects of axion-photon mixing on gamma-ray
  spectra from magnetized astrophysical sources,'' {\em Physical Review
  D—Particles, Fields, Gravitation, and Cosmology} {\bfseries 76} no.~12,
  (2007) 123011.

\bibitem{de2008axion}
A.~De~Angelis, O.~Mansutti, and M.~Roncadelli, ``Axion-like particles, cosmic
  magnetic fields and gamma-ray astrophysics,'' {\em Physics Letters B}
  {\bfseries 659} no.~5, (2008) 847--855.

\bibitem{abramowski2013constraints}
A.~Abramowski, F.~Acero, F.~Aharonian, F.~Ait~Benkhali, A.~Akhperjanian,
  E.~Ang{\"u}ner, G.~Anton, S.~Balenderan, A.~Balzer, A.~Barnacka, {\em
  et~al.}, ``Constraints on axionlike particles with hess from the irregularity
  of the pks 2155-304 energy spectrum,'' {\em Physical Review D—Particles,
  Fields, Gravitation, and Cosmology} {\bfseries 88} no.~10, (2013) 102003.

\bibitem{ajello2016search}
M.~Ajello, A.~Albert, B.~Anderson, L.~Baldini, G.~Barbiellini, D.~Bastieri,
  R.~Bellazzini, E.~Bissaldi, R.~Blandford, E.~Bloom, {\em et~al.}, ``Search
  for spectral irregularities due to photon--axionlike-particle oscillations
  with the fermi large area telescope,'' {\em Physical Review Letters}
  {\bfseries 116} no.~16, (2016) 161101.

\bibitem{marsh2017new}
M.~D. Marsh, H.~R. Russell, A.~C. Fabian, B.~R. McNamara, P.~Nulsen, and C.~S.
  Reynolds, ``A new bound on axion-like particles,'' {\em Journal of Cosmology
  and Astroparticle Physics} {\bfseries 2017} no.~12, (2017) 036.

\bibitem{zhang2018new}
C.~Zhang, Y.-F. Liang, S.~Li, N.-H. Liao, L.~Feng, Q.~Yuan, Y.-Z. Fan, and
  Z.-Z. Ren, ``New bounds on axionlike particles from the fermi large area
  telescope observation of pks 2155-304,'' {\em Physical Review D} {\bfseries
  97} no.~6, (2018) 063009.

\bibitem{reynolds2020astrophysical}
C.~S. Reynolds, M.~D. Marsh, H.~R. Russell, A.~C. Fabian, R.~Smith, F.~Tombesi,
  and S.~Veilleux, ``Astrophysical limits on very light axion-like particles
  from chandra grating spectroscopy of ngc 1275,'' {\em The Astrophysical
  Journal} {\bfseries 890} no.~1, (2020) 59.

\bibitem{freese1990natural}
K.~Freese, J.~A. Frieman, and A.~V. Olinto, ``Natural inflation with pseudo
  nambu-goldstone bosons,'' {\em Physical Review Letters} {\bfseries 65}
  no.~26, (1990) 3233.

\bibitem{kim2005completing}
J.~E. Kim, H.~P. Nilles, and M.~Peloso, ``Completing natural inflation,'' {\em
  Journal of Cosmology and Astroparticle Physics} {\bfseries 2005} no.~01,
  (2005) 005.

\bibitem{dimopoulos2008n}
S.~Dimopoulos, S.~Kachru, J.~McGreevy, and J.~G. Wacker, ``N-flation,'' {\em
  Journal of Cosmology and Astroparticle Physics} {\bfseries 2008} no.~08,
  (2008) 003.

\bibitem{preskill1983cosmology}
J.~Preskill, M.~B. Wise, and F.~Wilczek, ``Cosmology of the invisible axion,''
  {\em Physics Letters B} {\bfseries 120} no.~1-3, (1983) 127--132.

\bibitem{abbott1983cosmological}
L.~F. Abbott and P.~Sikivie, ``A cosmological bound on the invisible axion,''
  {\em Physics Letters B} {\bfseries 120} no.~1-3, (1983) 133--136.

\bibitem{dine1983not}
M.~Dine and W.~Fischler, ``The not-so-harmless axion,'' {\em Physics Letters B}
  {\bfseries 120} no.~1-3, (1983) 137--141.

\bibitem{hui2017ultralight}
L.~Hui, J.~P. Ostriker, S.~Tremaine, and E.~Witten, ``Ultralight scalars as
  cosmological dark matter,'' {\em Physical Review D} {\bfseries 95} no.~4,
  (2017) 043541.

\bibitem{chadha2022axion}
F.~Chadha-Day, J.~Ellis, and D.~J. Marsh, ``Axion dark matter: What is it and
  why now?,'' {\em Science advances} {\bfseries 8} no.~8, (2022) eabj3618.

\bibitem{dolan2022advancing}
M.~J. Dolan, F.~J. Hiskens, and R.~R. Volkas, ``Advancing globular cluster
  constraints on the axion-photon coupling,'' {\em Journal of Cosmology and
  Astroparticle Physics} {\bfseries 2022} no.~10, (2022) 096.

\bibitem{dessert2022upper}
C.~Dessert, D.~Dunsky, and B.~R. Safdi, ``Upper limit on the axion-photon
  coupling from magnetic white dwarf polarization,'' {\em Physical Review D}
  {\bfseries 105} no.~10, (2022) 103034.

\bibitem{Nomura:2022zyy}
K.~Nomura, K.~Saito, and J.~Soda, ``{Observing axions through photon ring
  dimming of black holes},''
  \href{http://dx.doi.org/10.1103/PhysRevD.107.123505}{{\em Phys. Rev. D}
  {\bfseries 107} no.~12, (2023) 123505},
  \href{http://arxiv.org/abs/2212.03020}{{\ttfamily arXiv:2212.03020
  [hep-ph]}}.

\bibitem{LIGOScientific:2016aoc}
{\bfseries LIGO Scientific, Virgo} Collaboration, B.~P. Abbott {\em et~al.},
  ``{Observation of Gravitational Waves from a Binary Black Hole Merger},''
  \href{http://dx.doi.org/10.1103/PhysRevLett.116.061102}{{\em Phys. Rev.
  Lett.} {\bfseries 116} no.~6, (2016) 061102},
  \href{http://arxiv.org/abs/1602.03837}{{\ttfamily arXiv:1602.03837 [gr-qc]}}.

\bibitem{LIGOScientific:2017vwq}
{\bfseries LIGO Scientific, Virgo} Collaboration, B.~P. Abbott {\em et~al.},
  ``{GW170817: Observation of Gravitational Waves from a Binary Neutron Star
  Inspiral},'' \href{http://dx.doi.org/10.1103/PhysRevLett.119.161101}{{\em
  Phys. Rev. Lett.} {\bfseries 119} no.~16, (2017) 161101},
  \href{http://arxiv.org/abs/1710.05832}{{\ttfamily arXiv:1710.05832 [gr-qc]}}.

\bibitem{Fish:2016jil}
{\bfseries Event Horizon Telescope} Collaboration, V.~L. Fish, K.~Akiyama,
  K.~L. Bouman, A.~A. Chael, M.~D. Johnson, S.~S. Doeleman, L.~Blackburn,
  J.~F.~C. Wardle, and W.~T. Freeman, ``{Observing\textemdash{}and
  Imaging\textemdash{}Active Galactic Nuclei with the Event Horizon
  Telescope},'' \href{http://dx.doi.org/10.3390/galaxies4040054}{{\em Galaxies}
  {\bfseries 4} no.~4, (2016) 54},
  \href{http://arxiv.org/abs/1607.03034}{{\ttfamily arXiv:1607.03034
  [astro-ph.IM]}}.

\bibitem{EventHorizonTelescope:2019dse}
{\bfseries Event Horizon Telescope} Collaboration, K.~Akiyama {\em et~al.},
  ``{First M87 Event Horizon Telescope Results. I. The Shadow of the
  Supermassive Black Hole},''
  \href{http://dx.doi.org/10.3847/2041-8213/ab0ec7}{{\em Astrophys. J. Lett.}
  {\bfseries 875} (2019) L1}, \href{http://arxiv.org/abs/1906.11238}{{\ttfamily
  arXiv:1906.11238 [astro-ph.GA]}}.

\bibitem{EventHorizonTelescope:2019uob}
{\bfseries Event Horizon Telescope} Collaboration, K.~Akiyama {\em et~al.},
  ``{First M87 Event Horizon Telescope Results. II. Array and
  Instrumentation},'' \href{http://dx.doi.org/10.3847/2041-8213/ab0c96}{{\em
  Astrophys. J. Lett.} {\bfseries 875} no.~1, (2019) L2},
  \href{http://arxiv.org/abs/1906.11239}{{\ttfamily arXiv:1906.11239
  [astro-ph.IM]}}.

\bibitem{EventHorizonTelescope:2019jan}
{\bfseries Event Horizon Telescope} Collaboration, K.~Akiyama {\em et~al.},
  ``{First M87 Event Horizon Telescope Results. III. Data Processing and
  Calibration},'' \href{http://dx.doi.org/10.3847/2041-8213/ab0c57}{{\em
  Astrophys. J. Lett.} {\bfseries 875} no.~1, (2019) L3},
  \href{http://arxiv.org/abs/1906.11240}{{\ttfamily arXiv:1906.11240
  [astro-ph.GA]}}.

\bibitem{EventHorizonTelescope:2019ths}
{\bfseries Event Horizon Telescope} Collaboration, K.~Akiyama {\em et~al.},
  ``{First M87 Event Horizon Telescope Results. IV. Imaging the Central
  Supermassive Black Hole},''
  \href{http://dx.doi.org/10.3847/2041-8213/ab0e85}{{\em Astrophys. J. Lett.}
  {\bfseries 875} no.~1, (2019) L4},
  \href{http://arxiv.org/abs/1906.11241}{{\ttfamily arXiv:1906.11241
  [astro-ph.GA]}}.

\bibitem{EventHorizonTelescope:2019pgp}
{\bfseries Event Horizon Telescope} Collaboration, K.~Akiyama {\em et~al.},
  ``{First M87 Event Horizon Telescope Results. V. Physical Origin of the
  Asymmetric Ring},'' \href{http://dx.doi.org/10.3847/2041-8213/ab0f43}{{\em
  Astrophys. J. Lett.} {\bfseries 875} no.~1, (2019) L5},
  \href{http://arxiv.org/abs/1906.11242}{{\ttfamily arXiv:1906.11242
  [astro-ph.GA]}}.

\bibitem{EventHorizonTelescope:2019ggy}
{\bfseries Event Horizon Telescope} Collaboration, K.~Akiyama {\em et~al.},
  ``{First M87 Event Horizon Telescope Results. VI. The Shadow and Mass of the
  Central Black Hole},'' \href{http://dx.doi.org/10.3847/2041-8213/ab1141}{{\em
  Astrophys. J. Lett.} {\bfseries 875} no.~1, (2019) L6},
  \href{http://arxiv.org/abs/1906.11243}{{\ttfamily arXiv:1906.11243
  [astro-ph.GA]}}.

\bibitem{Mayerson:2023wck}
D.~R. Mayerson and B.~Vercnocke, ``{Observational Opportunities for the
  Fuzzball Program},'' \href{http://arxiv.org/abs/2306.01565}{{\ttfamily
  arXiv:2306.01565 [hep-th]}}.

\bibitem{Dihingia:2024cch}
I.~K. Dihingia, A.~Uniyal, and Y.~Mizuno, ``{Distinguishability of a naked
  singularity from a black hole in dynamics and radiative signatures},''
  \href{http://arxiv.org/abs/2410.13406}{{\ttfamily arXiv:2410.13406
  [astro-ph.HE]}}.

\bibitem{Penrose:1969pc}
R.~Penrose, ``{Gravitational collapse: The role of general relativity},''
  \href{http://dx.doi.org/10.1023/A:1016578408204}{{\em Riv. Nuovo Cim.}
  {\bfseries 1} (1969) 252--276}.

\bibitem{Lora-Clavijo:2023ukh}
F.~D. Lora-Clavijo, G.~D. Prada-M\'endez, L.~M. Becerra, and E.~A.
  Becerra-Vergara, ``{The q-metric naked singularity: a viable explanation for
  the nature of the central object in the Milky Way},''
  \href{http://dx.doi.org/10.1088/1361-6382/ad0b9e}{{\em Class. Quant. Grav.}
  {\bfseries 40} no.~24, (2023) 245012},
  \href{http://arxiv.org/abs/2311.06653}{{\ttfamily arXiv:2311.06653 [gr-qc]}}.

\bibitem{Pal:2023wqg}
K.~Pal, K.~Pal, R.~Shaikh, and T.~Sarkar, ``{A rotating modified JNW spacetime
  as a Kerr black hole mimicker},''
  \href{http://dx.doi.org/10.1088/1475-7516/2023/11/060}{{\em JCAP} {\bfseries
  11} (2023) 060}, \href{http://arxiv.org/abs/2305.07518}{{\ttfamily
  arXiv:2305.07518 [gr-qc]}}.

\bibitem{Mishra:2023uxl}
R.~Mishra, R.~S.~S. Vieira, and W.~Klu\'zniak, ``{General upper limit on the
  electric charge of Sgr A* in the Reissner\textendash{}Nordstr\"om metric},''
  \href{http://dx.doi.org/10.1093/mnras/stae941}{{\em Mon. Not. Roy. Astron.
  Soc.} {\bfseries 530} no.~3, (2024) 3038--3042},
  \href{http://arxiv.org/abs/2304.04313}{{\ttfamily arXiv:2304.04313 [gr-qc]}}.

\bibitem{Deliyski:2023gik}
V.~Deliyski, G.~Gyulchev, P.~Nedkova, and S.~Yazadjiev, ``{Polarized image of
  equatorial emission in horizonless spacetimes: Naked singularities},''
  \href{http://dx.doi.org/10.1103/PhysRevD.108.104049}{{\em Phys. Rev. D}
  {\bfseries 108} no.~10, (2023) 104049},
  \href{http://arxiv.org/abs/2303.14756}{{\ttfamily arXiv:2303.14756 [gr-qc]}}.

\bibitem{Nguyen:2023clb}
B.~Nguyen, P.~Christian, and C.-k. Chan, ``{Shadow Geometry of Kerr Naked
  Singularities},'' \href{http://dx.doi.org/10.3847/1538-4357/ace697}{{\em
  Astrophys. J.} {\bfseries 954} no.~1, (2023) 78},
  \href{http://arxiv.org/abs/2302.08094}{{\ttfamily arXiv:2302.08094
  [astro-ph.HE]}}.

\bibitem{Roy:2023rjk}
S.~Roy, P.~Sarkar, S.~Sau, and S.~SenGupta, ``{Exploring axions through the
  photon ring of a spherically symmetric black hole},''
  \href{http://dx.doi.org/10.1088/1475-7516/2023/11/099}{{\em JCAP} {\bfseries
  11} (2023) 099}, \href{http://arxiv.org/abs/2310.05908}{{\ttfamily
  arXiv:2310.05908 [gr-qc]}}.

\bibitem{Chen:2024nua}
Y.~Chen, R.~Ding, Y.~Liu, Y.~Mizuno, J.~Shu, H.~Yu, and Y.~Zeng,
  ``{Illuminating Black Hole Shadow with Dark Matter Annihilation},''
  \href{http://arxiv.org/abs/2404.16673}{{\ttfamily arXiv:2404.16673
  [hep-ph]}}.

\bibitem{Chen:2022oad}
Y.~Chen, C.~Li, Y.~Mizuno, J.~Shu, X.~Xue, Q.~Yuan, Y.~Zhao, and Z.~Zhou,
  ``{Birefringence tomography for axion cloud},''
  \href{http://dx.doi.org/10.1088/1475-7516/2022/09/073}{{\em JCAP} {\bfseries
  09} (2022) 073}, \href{http://arxiv.org/abs/2208.05724}{{\ttfamily
  arXiv:2208.05724 [hep-ph]}}.

\bibitem{Raffelt:1987im}
G.~Raffelt and L.~Stodolsky, ``{Mixing of the Photon with Low Mass
  Particles},'' \href{http://dx.doi.org/10.1103/PhysRevD.37.1237}{{\em Phys.
  Rev. D} {\bfseries 37} (1988) 1237}.

\bibitem{Hochmuth:2007hk}
K.~A. Hochmuth and G.~Sigl, ``{Effects of Axion-Photon Mixing on Gamma-Ray
  Spectra from Magnetized Astrophysical Sources},''
  \href{http://dx.doi.org/10.1103/PhysRevD.76.123011}{{\em Phys. Rev. D}
  {\bfseries 76} (2007) 123011},
  \href{http://arxiv.org/abs/0708.1144}{{\ttfamily arXiv:0708.1144
  [astro-ph]}}.

\bibitem{Masaki:2017aea}
E.~Masaki, A.~Aoki, and J.~Soda, ``{Photon-Axion Conversion, Magnetic Field
  Configuration, and Polarization of Photons},''
  \href{http://dx.doi.org/10.1103/PhysRevD.96.043519}{{\em Phys. Rev. D}
  {\bfseries 96} no.~4, (2017) 043519},
  \href{http://arxiv.org/abs/1702.08843}{{\ttfamily arXiv:1702.08843
  [astro-ph.CO]}}.

\bibitem{EventHorizonTelescope:2021bee}
{\bfseries Event Horizon Telescope} Collaboration, K.~Akiyama {\em et~al.},
  ``{First M87 Event Horizon Telescope Results. VII. Polarization of the
  Ring},'' \href{http://dx.doi.org/10.3847/2041-8213/abe71d}{{\em Astrophys. J.
  Lett.} {\bfseries 910} no.~1, (2021) L12},
  \href{http://arxiv.org/abs/2105.01169}{{\ttfamily arXiv:2105.01169
  [astro-ph.HE]}}.

\bibitem{EventHorizonTelescope:2021srq}
{\bfseries Event Horizon Telescope} Collaboration, K.~Akiyama {\em et~al.},
  ``{First M87 Event Horizon Telescope Results. VIII. Magnetic Field Structure
  near The Event Horizon},''
  \href{http://dx.doi.org/10.3847/2041-8213/abe4de}{{\em Astrophys. J. Lett.}
  {\bfseries 910} no.~1, (2021) L13},
  \href{http://arxiv.org/abs/2105.01173}{{\ttfamily arXiv:2105.01173
  [astro-ph.HE]}}.

\bibitem{Gralla:2019xty}
S.~E. Gralla, D.~E. Holz, and R.~M. Wald, ``{Black Hole Shadows, Photon Rings,
  and Lensing Rings},''
  \href{http://dx.doi.org/10.1103/PhysRevD.100.024018}{{\em Phys. Rev. D}
  {\bfseries 100} no.~2, (2019) 024018},
  \href{http://arxiv.org/abs/1906.00873}{{\ttfamily arXiv:1906.00873
  [astro-ph.HE]}}.

\bibitem{PhysRevLett.20.878}
A.~I. Janis, E.~T. Newman, and J.~Winicour, ``Reality of the schwarzschild
  singularity,'' \href{http://dx.doi.org/10.1103/PhysRevLett.20.878}{{\em Phys.
  Rev. Lett.} {\bfseries 20} (Apr, 1968) 878--880}.
  \url{https://link.aps.org/doi/10.1103/PhysRevLett.20.878}.

\bibitem{Fisher:1948yn}
I.~Z. Fisher, ``{Scalar mesostatic field with regard for gravitational
  effects},'' {\em Zh. Eksp. Teor. Fiz.} {\bfseries 18} (1948) 636--640,
  \href{http://arxiv.org/abs/gr-qc/9911008}{{\ttfamily arXiv:gr-qc/9911008}}.

\bibitem{10.1093/mnras/215.4.575}
M.~Y. Khlopov, B.~A. Malomed, and Y.~B. Zeldovich, ``{Gravitational instability
  of scalar fields and formation of primordial black holes},''
  \href{http://dx.doi.org/10.1093/mnras/215.4.575}{{\em Monthly Notices of the
  Royal Astronomical Society} {\bfseries 215} no.~4, (08, 1985) 575--589},
  \href{http://arxiv.org/abs/https://academic.oup.com/mnras/article-pdf/215/4/575/4082842/mnras215-0575.pdf}{{\ttfamily
  https://academic.oup.com/mnras/article-pdf/215/4/575/4082842/mnras215-0575.pdf}}.
  \url{https://doi.org/10.1093/mnras/215.4.575}.

\bibitem{Virbhadra:1997ie}
K.~S. Virbhadra, ``{Janis-Newman-Winicour and Wyman solutions are the same},''
  \href{http://dx.doi.org/10.1142/S0217751X97002577}{{\em Int. J. Mod. Phys. A}
  {\bfseries 12} (1997) 4831--4836},
  \href{http://arxiv.org/abs/gr-qc/9701021}{{\ttfamily arXiv:gr-qc/9701021}}.

\bibitem{PhysRevD.24.839}
M.~Wyman, ``Static spherically symmetric scalar fields in general relativity,''
  \href{http://dx.doi.org/10.1103/PhysRevD.24.839}{{\em Phys. Rev. D}
  {\bfseries 24} (Aug, 1981) 839--841}.
  \url{https://link.aps.org/doi/10.1103/PhysRevD.24.839}.

\bibitem{Gyulchev:2008ff}
G.~N. Gyulchev and S.~S. Yazadjiev, ``{Gravitational Lensing by Rotating Naked
  Singularities},'' \href{http://dx.doi.org/10.1103/PhysRevD.78.083004}{{\em
  Phys. Rev. D} {\bfseries 78} (2008) 083004},
  \href{http://arxiv.org/abs/0806.3289}{{\ttfamily arXiv:0806.3289 [gr-qc]}}.

\bibitem{PhysRevD.65.103004}
K.~S. Virbhadra and G.~F.~R. Ellis, ``Gravitational lensing by naked
  singularities,'' \href{http://dx.doi.org/10.1103/PhysRevD.65.103004}{{\em
  Phys. Rev. D} {\bfseries 65} (May, 2002) 103004}.
  \url{https://link.aps.org/doi/10.1103/PhysRevD.65.103004}.

\bibitem{Virbhadra:1998dy}
K.~S. Virbhadra, D.~Narasimha, and S.~M. Chitre, ``{Role of the scalar field in
  gravitational lensing},'' {\em Astron. Astrophys.} {\bfseries 337} (1998)
  1--8, \href{http://arxiv.org/abs/astro-ph/9801174}{{\ttfamily
  arXiv:astro-ph/9801174}}.

\bibitem{Virbhadra:2007kw}
K.~S. Virbhadra and C.~R. Keeton, ``{Time delay and magnification centroid due
  to gravitational lensing by black holes and naked singularities},''
  \href{http://dx.doi.org/10.1103/PhysRevD.77.124014}{{\em Phys. Rev. D}
  {\bfseries 77} (2008) 124014},
  \href{http://arxiv.org/abs/0710.2333}{{\ttfamily arXiv:0710.2333 [gr-qc]}}.

\bibitem{Yang:2015hwf}
L.~Yang and Z.~Li, ``{Shadow of a dressed black hole and determination of spin
  and viewing angle},'' \href{http://dx.doi.org/10.1142/S0218271816500267}{{\em
  Int. J. Mod. Phys. D} {\bfseries 25} no.~02, (2015) 1650026},
  \href{http://arxiv.org/abs/1511.00086}{{\ttfamily arXiv:1511.00086
  [astro-ph.HE]}}.

\bibitem{Takahashi:2004xh}
R.~Takahashi, ``{Shapes and positions of black hole shadows in accretion disks
  and spin parameters of black holes},''
  \href{http://dx.doi.org/10.1086/422403}{{\em J. Korean Phys. Soc.} {\bfseries
  45} (2004) S1808--S1812},
  \href{http://arxiv.org/abs/astro-ph/0405099}{{\ttfamily
  arXiv:astro-ph/0405099}}.

\bibitem{PhysRevD.100.024055}
G.~Gyulchev, P.~Nedkova, T.~Vetsov, and S.~Yazadjiev, ``Image of the
  janis-newman-winicour naked singularity with a thin accretion disk,''
  \href{http://dx.doi.org/10.1103/PhysRevD.100.024055}{{\em Phys. Rev. D}
  {\bfseries 100} (Jul, 2019) 024055}.
  \url{https://link.aps.org/doi/10.1103/PhysRevD.100.024055}.

\bibitem{Chowdhury:2011aa}
A.~N. Chowdhury, M.~Patil, D.~Malafarina, and P.~S. Joshi, ``{Circular
  geodesics and accretion disks in Janis-Newman-Winicour and Gamma metric},''
  \href{http://dx.doi.org/10.1103/PhysRevD.85.104031}{{\em Phys. Rev. D}
  {\bfseries 85} (2012) 104031},
  \href{http://arxiv.org/abs/1112.2522}{{\ttfamily arXiv:1112.2522 [gr-qc]}}.

\bibitem{Sau:2020xau}
S.~Sau, I.~Banerjee, and S.~SenGupta, ``{Imprints of the Janis-Newman-Winicour
  spacetime on observations related to shadow and accretion},''
  \href{http://dx.doi.org/10.1103/PhysRevD.102.064027}{{\em Phys. Rev. D}
  {\bfseries 102} no.~6, (2020) 064027},
  \href{http://arxiv.org/abs/2004.02840}{{\ttfamily arXiv:2004.02840 [gr-qc]}}.

\bibitem{Chen:2023uuy}
D.~Chen, Y.~Chen, P.~Wang, T.~Wu, and H.~Wu, ``{Gravitational lensing by
  transparent Janis\textendash{}Newman\textendash{}Winicour naked
  singularities},''
  \href{http://dx.doi.org/10.1140/epjc/s10052-024-12950-z}{{\em Eur. Phys. J.
  C} {\bfseries 84} no.~6, (2024) 584},
  \href{http://arxiv.org/abs/2309.00905}{{\ttfamily arXiv:2309.00905 [gr-qc]}}.

\bibitem{Stashko:2022dtx}
O.~S. Stashko and V.~I. Zhdanov, ``{Circular orbits of test particles
  interacting with massless linear scalar field of the naked singularity},''
  \href{http://dx.doi.org/10.1103/PhysRevD.106.104049}{{\em Phys. Rev. D}
  {\bfseries 106} no.~10, (2022) 104049},
  \href{http://arxiv.org/abs/2209.00160}{{\ttfamily arXiv:2209.00160 [gr-qc]}}.

\bibitem{Patel:2022vlu}
V.~Patel, D.~Tahelyani, A.~B. Joshi, D.~Dey, and P.~S. Joshi, ``{Light
  trajectory and shadow shape in the rotating naked singularity},''
  \href{http://dx.doi.org/10.1140/epjc/s10052-022-10638-w}{{\em Eur. Phys. J.
  C} {\bfseries 82} no.~9, (2022) 798},
  \href{http://arxiv.org/abs/2206.06750}{{\ttfamily arXiv:2206.06750 [gr-qc]}}.

\bibitem{Vagnozzi:2022moj}
S.~Vagnozzi {\em et~al.}, ``{Horizon-scale tests of gravity theories and
  fundamental physics from the Event Horizon Telescope image of Sagittarius
  A},'' \href{http://dx.doi.org/10.1088/1361-6382/acd97b}{{\em Class. Quant.
  Grav.} {\bfseries 40} no.~16, (2023) 165007},
  \href{http://arxiv.org/abs/2205.07787}{{\ttfamily arXiv:2205.07787 [gr-qc]}}.

\bibitem{Solanki:2021mkt}
D.~N. Solanki, P.~Bambhaniya, D.~Dey, P.~S. Joshi, and K.~N. Pathak, ``{Shadows
  and precession of orbits in rotating
  Janis\textendash{}Newman\textendash{}Winicour spacetime},''
  \href{http://dx.doi.org/10.1140/epjc/s10052-022-10045-1}{{\em Eur. Phys. J.
  C} {\bfseries 82} no.~1, (2022) 77},
  \href{http://arxiv.org/abs/2109.14937}{{\ttfamily arXiv:2109.14937 [gr-qc]}}.

\bibitem{Joshi:2020tlq}
A.~B. Joshi, D.~Dey, P.~S. Joshi, and P.~Bambhaniya, ``{Shadow of a Naked
  Singularity without Photon Sphere},''
  \href{http://dx.doi.org/10.1103/PhysRevD.102.024022}{{\em Phys. Rev. D}
  {\bfseries 102} no.~2, (2020) 024022},
  \href{http://arxiv.org/abs/2004.06525}{{\ttfamily arXiv:2004.06525 [gr-qc]}}.

\bibitem{Deliyski:2024wmt}
V.~Deliyski, G.~Gyulchev, P.~Nedkova, and S.~Yazadjiev, ``{Observing naked
  singularities by the present and next-generation Event Horizon Telescope},''
  \href{http://arxiv.org/abs/2401.14092}{{\ttfamily arXiv:2401.14092 [gr-qc]}}.

\bibitem{Masaki:2019ggg}
E.~Masaki, A.~Aoki, and J.~Soda, ``{Stability of Axion Dark Matter-Photon
  Conversion},'' \href{http://dx.doi.org/10.1103/PhysRevD.101.043505}{{\em
  Phys. Rev. D} {\bfseries 101} no.~4, (2020) 043505},
  \href{http://arxiv.org/abs/1909.11470}{{\ttfamily arXiv:1909.11470
  [hep-ph]}}.

\bibitem{Dror:2021nyr}
J.~A. Dror, H.~Murayama, and N.~L. Rodd, ``{Cosmic axion background},''
  \href{http://dx.doi.org/10.1103/PhysRevD.103.115004}{{\em Phys. Rev. D}
  {\bfseries 103} no.~11, (2021) 115004},
  \href{http://arxiv.org/abs/2101.09287}{{\ttfamily arXiv:2101.09287
  [hep-ph]}}. [Erratum: Phys.Rev.D 106, 119902 (2022)].

\bibitem{Brito:2015oca}
R.~Brito, V.~Cardoso, and P.~Pani, ``{Superradiance}: {New Frontiers in Black
  Hole Physics},'' \href{http://dx.doi.org/10.1007/978-3-319-19000-6}{{\em
  Lect. Notes Phys.} {\bfseries 906} (2015) pp.1--237},
  \href{http://arxiv.org/abs/1501.06570}{{\ttfamily arXiv:1501.06570 [gr-qc]}}.

\bibitem{Day:2019bbh}
F.~V. Day and J.~I. McDonald, ``{Axion superradiance in rotating neutron
  stars},'' \href{http://dx.doi.org/10.1088/1475-7516/2019/10/051}{{\em JCAP}
  {\bfseries 10} (2019) 051}, \href{http://arxiv.org/abs/1904.08341}{{\ttfamily
  arXiv:1904.08341 [hep-ph]}}.

\bibitem{Pal:2022cxb}
K.~Pal, K.~Pal, P.~Roy, and T.~Sarkar, ``{Regularizing the JNW and JMN naked
  singularities},''
  \href{http://dx.doi.org/10.1140/epjc/s10052-023-11558-z}{{\em Eur. Phys. J.
  C} {\bfseries 83} no.~5, (2023) 397},
  \href{http://arxiv.org/abs/2206.11764}{{\ttfamily arXiv:2206.11764 [gr-qc]}}.

\bibitem{10.1093/mnras/sty2624}
R.~Shaikh, P.~Kocherlakota, R.~Narayan, and P.~S. Joshi, ``{Shadows of
  spherically symmetric black holes and naked singularities},''
  \href{http://dx.doi.org/10.1093/mnras/sty2624}{{\em Monthly Notices of the
  Royal Astronomical Society} {\bfseries 482} no.~1, (10, 2018) 52--64},
  \href{http://arxiv.org/abs/https://academic.oup.com/mnras/article-pdf/482/1/52/26145030/sty2624.pdf}{{\ttfamily
  https://academic.oup.com/mnras/article-pdf/482/1/52/26145030/sty2624.pdf}}.
  \url{https://doi.org/10.1093/mnras/sty2624}.

\bibitem{Narayan:2019imo}
R.~Narayan, M.~D. Johnson, and C.~F. Gammie, ``{The Shadow of a Spherically
  Accreting Black Hole},''
  \href{http://dx.doi.org/10.3847/2041-8213/ab518c}{{\em Astrophys. J. Lett.}
  {\bfseries 885} no.~2, (2019) L33},
  \href{http://arxiv.org/abs/1910.02957}{{\ttfamily arXiv:1910.02957
  [astro-ph.HE]}}.

\bibitem{Bambi:2013nla}
C.~Bambi, ``{Can the supermassive objects at the centers of galaxies be
  traversable wormholes? The first test of strong gravity for mm/sub-mm very
  long baseline interferometry facilities},''
  \href{http://dx.doi.org/10.1103/PhysRevD.87.107501}{{\em Phys. Rev. D}
  {\bfseries 87} (2013) 107501},
  \href{http://arxiv.org/abs/1304.5691}{{\ttfamily arXiv:1304.5691 [gr-qc]}}.

\bibitem{rybicki2004radiative}
G.~B. Rybicki and A.~P. Lightman, ``Radiative processes in astrophysics.
  originally published,'' 2004.

\bibitem{Quataert:2002xn}
E.~Quataert, ``{A thermal bremsstrahlung model for the quiescent x-ray emission
  from sagittarius a*},'' \href{http://dx.doi.org/10.1086/341425}{{\em
  Astrophys. J.} {\bfseries 575} (2002) 855--859},
  \href{http://arxiv.org/abs/astro-ph/0201395}{{\ttfamily
  arXiv:astro-ph/0201395}}.

\bibitem{Yuan:2014gma}
F.~Yuan and R.~Narayan, ``{Hot Accretion Flows Around Black Holes},''
  \href{http://dx.doi.org/10.1146/annurev-astro-082812-141003}{{\em Ann. Rev.
  Astron. Astrophys.} {\bfseries 52} (2014) 529--588},
  \href{http://arxiv.org/abs/1401.0586}{{\ttfamily arXiv:1401.0586
  [astro-ph.HE]}}.

\end{thebibliography}\endgroup
\bibliographystyle{./utphys1}


\end{document}